\documentclass{article}
\usepackage{epsfig}
\usepackage{changebar}

\newcommand{\Rr}{{\rm I\!R}}

\newcommand{\N}{{\rm I\!N}}

\newcommand{\half}{\frac{1}{2}}

\newcommand{\C}{{\mathchoice {\setbox0=\hbox{$\displaystyle\rm C$}\hbox{\hbox
to0pt{\kern0.4\wd0\vrule  height0.9\ht0\hss}\box0}}
{\setbox0=\hbox{$\textstyle\rm C$}\hbox{\hbox
to0pt{\kern0.4\wd0\vrule  height0.9\ht0\hss}\box0}}
{\setbox0=\hbox{$\scriptstyle\rm C$}\hbox{\hbox
to0pt{\kern0.4\wd0\vrule  height0.9\ht0\hss}\box0}}
{\setbox0=\hbox{$\scriptscriptstyle\rm C$}\hbox{\hbox
to0pt{\kern0.4\wd0\vrule  height0.9\ht0\hss}\box0}}}} %end of macro for \C

\newcommand{\Z}{{{\mathchoice  {\hbox{$\textstyle  Z\kern-0.4em  Z$}}
{\hbox{$\textstyle  Z\kern-0.4em  Z$}}
{\hbox{$\scriptstyle  Z\kern-0.3em  Z$}}
{\hbox{$\scriptscriptstyle  Z\kern-0.2em  Z$}}}}}

\newcommand{\Q}{{\mathchoice   {\setbox0=\hbox{$\displaystyle\rm
Q$}\hbox{\raise
0.15\ht0\hbox  to0pt{\kern0.4\wd0\vrule  height0.8\ht0\hss}\box0}}
{\setbox0=\hbox{$\textstyle\rm  Q$}\hbox{\raise
0.15\ht0\hbox  to0pt{\kern0.4\wd0\vrule  height0.8\ht0\hss}\box0}}
{\setbox0=\hbox{$\scriptstyle\rm  Q$}\hbox{\raise
0.15\ht0\hbox  to0pt{\kern0.4\wd0\vrule  height0.7\ht0\hss}\box0}}
{\setbox0=\hbox{$\scriptscriptstyle\rm  Q$}\hbox{\raise
0.15\ht0\hbox  to0pt{\kern0.4\wd0\vrule  height0.7\ht0\hss}\box0}}}}

\newcommand{\qed}{\ \ \rule{1ex}{1ex}}

\begin{document}

\begin{title}{ The inverse spectral problem for surfaces of
revolution} \end{title}

\begin{author}{ Steve Zelditch\\Johns Hopkins University, Baltimore, Maryland 
21218\\zel@math.jhu.edu}\end{author}

\begin{thanks}{Research 
partially supported by NSF grants
\#DMS-9404637 and \#DMS-9703775.}\end{thanks}

\maketitle

\addtolength{\baselineskip}{1pt} 

\setcounter{page}{-1}
\setcounter{section}{-1}
\begin{abstract}  We prove that isospectral simple analytic  surfaces of revolution
are isometric.

\end{abstract}

\section{Introduction} This article is concerned with the inverse spectral problem
for metrics of revolution on $S^2$. We will assume that our metrics are real
analytic and belong to a class   ${\cal R}^*$ of rotationally invariant metrics which
are of `simple type' and which satisfy some generic non-degeneracy  conditions (see
Definition (0.1)).  In particular, we will assume they satisfy the generalized
`simple length spectrum' condition that the length functional on the loop space
is a clean Bott-Morse function which takes on distinct values on distinct components
of its critical set (up to orientation).  Denoting by $Spec(S^2,g)$ the spectrum of
the Laplacian
$\Delta_g$, our main result is the following:
\bigskip

\noindent{\bf  Theorem I }~~~{\it Spec: ${\cal R}^* \rightarrow \Rr^{+ \N}$ is 1-1.}
\bigskip
 
Thus, if $(S^2,g), (S^2,h)$ are isospectral 
surfaces of revolution in ${\cal R}^*$, then $g$ is isometric to $h$.  
 It would be very
desirable to strengthen this result by removing the assumption that $h \in {\cal
R}^*$,  thereby showing that metrics in ${\cal R}^*$ are spectrally determined
within the entire class of analytic metrics on $S^2$ with simple length spectra.
The only metric on $S^2$ presently known to be spectrally determined in this sense is
the standard one (which is known to be spectrally determined among all $C^{\infty}$
metrics).  A metric $h$ satisfying $Spec(S^2, h) = Spec(S^2, g)$ for some
$g \in {\cal R}^*$ must have many properties in common with a surface of revolution
of simple type; it would be interesting to explore whether it must necessarily
be one.

Let us now  be more precise about the hypotheses. First, we will assume that there
is an effective action of $S^1$ by isometries of
$(S^2, g)$.  The two fixed points will be denoted $N,S$ and $(r,
\theta)$ will denote geodesic polar coordinates centered at $N$, with $\theta = 0$
some fixed meridian
$\gamma_M$ from $N$ to $S$. The metric may then be written in the form
$g = dr^2 + a(r)^2 d\theta^2$ where $a: [0,L] \rightarrow \Rr^+$
is defined by $a(r) = \frac{1}{2\pi} |S_r(N)|$, with $|S_r(N)|$  the length of
 the distance circle of radius $r$ centered at $N$.  For any smooth surface of
revolution, the function $a$ satisfies 
$a^{(2p)}(0) = a^{(2p)}(L) = 0, a'(0) = 1, a'(L) = -1$ and two such surfaces $
(S^2, g_i)$ $(i=1,2)$ are isometric if and only if $L_1 = L_2$ and $a_1(r) = a_2(r)$
or
$a_1(r) = a_2(L-r).$ We will then assume that the metrics belong to the following
class
 ${\cal R}$ of {\it simple analytic} surfaces of revolution [Bl]:
\bigskip

\noindent{\bf (0.1) Definition}~~~{\it  ${\cal R}$ is the moduli space of metrics
of revolution $(S^2,g)$ with the properties:\\
\noindent(i) $g$ (equivalently $a$) is real analytic;\\
\noindent(ii) $a$ has precisely one non-degenerate critical point $r_o \in (0, L)$,
with $a''(r_o) < 0$,
corresponding to an `equatorial geodesic' $\gamma_E$;\\
\noindent(iii)  the (non-linear) Poincare map ${\cal P}_{\gamma_E}$ for
$\gamma_E$ is of twist type (cf. \S 1). \\
 We denote by ${\cal R}^* \subset {\cal R}$  the  subset of metrics 
with  `simple length spectra' in the sense above.}
\bigskip

Regarding `simple length spectra,'  we recall  that    the closed
geodesics of a surface of revolution come in one-parameter families of a common
length, filling out invariant torii ${\cal T}$
 for the geodesic flow.  The canonical involution $\sigma(x,\xi) = (x, - \xi)$
takes  ${\cal T}$ to its `time reversal' $- {\cal T}$, and takes the closed
geodesics of ${\cal T}$ to their reversals on $- {\cal T}$.  A closed geodesic
and its reversal have the same length, so the length spectrum is automatically double
except for the length $2L$ of the torus of meridians, which is $\sigma$-invariant.
The simple length spectrum hypothesis is that up to time reversal, the common lengths
of the closed geodesics on distinct torii are  distinct   (cf. Definition 1.2.2).  In
fact, it would be sufficient for the proof that  the length $2L$ of the `meridian'
closed geodesics is not the length of  closed geodesics on any other torus.
In any case, it is not hard to show that ${\cal R}^*$
is residual in ${\cal R}$ (cf. Proposition 1.2.4).

The condition (iii) actually appears in the proof in the following way:
 the quadratic coefficient `$\alpha := h''(0)$' of the classical Birkhoff normal form
of the metric $|\xi|_g$ at the torus of meridian geodesics must be non-vanishing (see
Definition (1.4.5) for the precise meaning of $h(\xi)$). This condition is used 
 in Proposition (4.1.2) and Corollary (4.1.3) to
evaluate  the wave invariants for the meridian torus.

As will be explained
further below, the main purpose of the   non-degeneracy and simplicity
conditions is to insure that there are global action-angle variables for the geodesic
flow. These conditions rule out several types of
surfaces of revolution:  First, they rule out Zoll surfaces
of revolution, 
which are degenerate in every possible sense. It is indeed unknown at this time
whether real analytic Zoll surfaces of revolution are determined by their spectra. 
They also rule out  `peanuts of revolution' (which have hyperbolic waists) and
other natural rotational surfaces such as Liouville torii.

 To our knowledge, the strongest prior result on the inverse spectral problem
for surfaces of revolution is that of Bruning-Heintze [B.H]: smooth surfaces
of revolution with a mirror symmetry thru the $x-y$ plane are spectrally determined
among metrics of this kind. There are also a number of proofs that a surface
of revolution is determined by the joint spectrum of $\Delta$ and of $\partial
/\partial \theta$, the generator of the rotational symmetry [Kac][B][Gur]. 
 
 The method-of-proof of Bruning-Heintze was based on the observation that
the invariant spectrum can be heard from the entire spectrum.  Hence by separating
variables the problem can be reduced to the inverse spectral problem for 1D
even singular Sturm-Liouville operators, which was solved by Gelfand-Levitan and
Marchenko.

Our proof of the Main Theorem is based on different kind of method, and the inverse
result presented here is hopefully just one illustration of it. We begin with the
facts that simple surfaces of revolution are completely integrable  on both the
classical and quantum levels and that the Laplacian has a global quantum normal form
in terms of action operators.  We then study the trace of the wave group and prove
that from its singularity expansion we can reconstruct the global
quantum  normal form.  Finally we  show that this normal form determines the
metric.

This approach is suggested by the recent inverse result of Guillemin [G.1], which
shows that the microlocal normal form of $\Delta$ around each non-degenerate elliptic
closed geodesic can be determined  from the wave invariants (see also
[Z.1,2]).  However, there is but one non-degenerate  closed geodesic on a simple
surface of revolution, so the direct application of this inverse result does not take
full advantage of the situation.  Rather, it is natural to start from the fact
that the wave group is completely integrable in the following 
strong sense:  namely, it commutes with an effective action of the torus $S^1
\times S^1$ by Fourier Integral operators on $L^2(S^2)$.  That is, there exist
global {\it action operators}
$\hat{I}_1, \hat{I}_2$ and a polyhomogeneous symbol $\hat{H}$ of degree 1 on
$\Rr^2 - 0$ such that $\sqrt{\Delta} = \hat{H}(\hat{I_1}, \hat{I_2})$.
This is the global quantum Birkhoff normal form alluded to above. Our principal
tool is the following inverse result:
\bigskip

\noindent{\bf Main Lemma}~~~{ \it The wave trace invariants of $(S^2,g)$ with $g\in
{\cal R}^*$  determine the quantum normal form $\hat{H}.$}
\bigskip

This Lemma does not actually require that $(S^2,g)$ be a surface of revolution,
but only that the geodesic flow is toric integrable, i.e. commutes with an effective 
Hamiltonian torus action.
It  immediately implies that
the principal symbol $H(I_1, I_2)$ of $\hat{H}(\hat{I}_1, \hat{I}_2)$ is a spectral
invariant. Since $H(I_1, I_2)$ is essentially a global Birkhoff normal form for the
metric, the wave invariants determine the symplectic equivalence of the geodesic flow.
Thus we have:
\bigskip

\noindent{\bf Corollary 1}~~~{\it  From the wave trace invariants of $(S^2,g)$ with
$g\in {\cal R}^*$ we can determine the symplectic equivalence class of $G^t$.}
\bigskip

 Corollary 1 does not however  determine the isometry class of a general $g \in
{\cal R}^*$: As will be discussed in \S 1 (see also [C.K]), simple surfaces of
revolution are not symplectically rigid unless they are mirror symmetric. 
Otherwise put, recovery of the classical Birkhoff normal form only 
determines the {\it even part} of the metric in the following sense:

\bigskip

\noindent{\bf Corollary 2}~~~{\it A metric $g \in {\cal R}$ may be written in the form
$g = [f(cos u)]^2 du^2 + [sin u]^2 d\theta^2$ (\S 1, [Besse]). From the 
wave invariants one
can determine the {\em even part} of $f$.  In particular if $g$ is mirror symmetric, one can
determine $g$ among mirror symmetric metrics in ${\cal R}^*$ from its spectrum.}
\bigskip

It is interesting to note that symplectic 
 rigidity in the mirror-symmetric case 
gives a new and self-contained  proof of the Bruning-Heintze result, without
the use of Marchenko's inverse spectral theorem for singular
 Sturm Liouville operators.   Although this
result is superceded by the   Theorem, it may have some
future relevance to other inverse problems.

To complete the proof of the Theorem, we therefore have to study the subprincipal
terms in $\hat{H}$.  The result is:
\bigskip

\noindent{\bf Final Lemma }~~~{\it From $\hat{H}$ one can determine the isometry
class of $g$.}
\bigskip

It is in this last Lemma that we  use in full that $(S^2,g)$ is a surface of
revolution rather than just a surface with toric integrable geodesic flow.
We also use in full that $\hat{H}$ is a global quantum normal form rather than
a microlocal one at $\gamma_E$.  In subsequent work we will investigate the
analogues of the Final Lemma for the microlocal normal form at $\gamma_E$, for more
general toric integrable metrics and for metrics isospectral to toric integrable ones.

To close this introduction, we discuss some background and some open problems
connected with this work.  

The principal motivation for studying the inverse spectral problem for  surfaces
of revolution is its simplicity.  There are to date very few inverse results which
determine a metric from its Laplace spectrum within the entire class of metrics, or
even within concrete infinite dimensional families. To our knowledge, only the
standard $S^n$ for $n \leq 6$ and flat 2-torii are known to be
spectrally determined.
Hence it is desirable to have a simple model of how an inverse result might go. 

A second motivation is a somewhat loose analogy between surfaces of revolution
and planar domains.
Namely, in both cases the unknown is a function of one variable (the profile curve,
resp. the boundary) which completely determines
the first return times and angles of geodesics emanating from a transversal.  That is,
 paths of billiard trajectories (broken  geodesics) on a bounded planar domain are 
determined from  collisions  with the boundary, while  paths of  geodesics on
 surfaces of revolution are determined from  collisions with a meridian (or with the
equator).  In analytic cases, it is plausible that the unknown
function may be determined in large part by the spectrum of first return times from
the local transversal.  In the case of an analytic surface of revolution, this
length spectrum determines the corresonding Birkhoff normal form 
 and hence by Corollary 2 it determines the even part of the profile curve.   
Similarly, in the case of an analytic plane domain,   it is proved in 
[CV.2] that the even part of the boundary of an analytic domain may be determined from
the Birkhoff normal form of the billiard map at a bouncing ball orbit.  
Hence, there is a similarity in the relation between the unknown function and the
local classical Birkhoff normal forms.   It is interesting to observe in this context
that the rigidity result of Colin de Verdiere [CV.2] is quite analogous to
the inverse result of Bruning-Heintze.  

Some immediate open problems:
First, there is the symplectic conjugacy problem mentioned above.  Second,
can one relax analyticity to smoothness in the Theorem above?  This is likely
to follow from a more intensive analysis of the wave invariants.  Third, can one
extend it to other `non-simple' types of surfaces of revolution?  The main obstacle
is that one will generally not have global action-angle variables or  global quantum
normal forms.  What about completely integrable systems in higher dimensions?   
 
Finally, we would like to  thank  D.Kosygin for several helpful
conversations on this paper, and   B.Kleiner for  giving us up to date information
about the status of the conjugacy problem for geodesic flows on surfaces of
revolution.

\section{Classical dynamics}

\subsection{Global action-angle variables}

From a geometric (or dynamical) point of view, the principal virtue of metrics
in ${\cal R}$ is described by the following:
\bigskip

\noindent{\bf (1.1.1) Proposition}~~~{\it Suppose $g$ is a real analytic metric of
revolution on
$S^2$ such that $a$ has precisely one non-degenerate critical point at some
$r_o \in (0, L).$  Then the Hamiltonian $|\xi|_g := \sqrt{\sum g^{ij}\xi_i \xi_j}$ 
on $T^*S^2$  is  completely integrable and possesses global real analytic action-angle
variables.}
\bigskip

\noindent{\bf Proof}: The complete integrability of $|\xi|_g$ (i.e. of the
geodesic flow) is classical, and follows from the existence of the Clairaut
integral $p_{\theta}(v):= \langle v, \frac{\partial}{\partial \theta}\rangle$.  Since
 the Poisson bracket $\{p_{\theta}, |\xi|_g\} = 0$, the geodesics
are constrained to lie on the level sets of $p_{\theta}$; and  since
both $|\xi|_g$ and $p_{\theta}$ are homogeneous of degree one, the behaviour
of the geodesic flow is determined by its restriction to $S^*_gS^2 = \{|\xi|_g =
1\}$.
With the assumption on
$a$, the level sets are compact and the only critical level is that of the 
equatorial geodesics $\gamma_E^{\pm} \subset S^*_gS^2$ (traversed with either
orientation).  The other level sets are well-known to consist of two-dimensional
torii.  (Had we allowed the existence of at least two critical points in $a$,
there would exist a saddle level, i.e. an embedded non-compact
cylinder).

The existence  of global action-angle variables follows from the general results
of [D][G.S] and have been constructed explicitly for simple surfaces of revolution
in [CV.1].  The general formula is as follows: Let 
$$P = (|\xi|_g, p_{\theta}): T^*S^2 \rightarrow B:= \{(b_1, b_2) :
|b_2| \leq a(r_o)b_1\} \subset
\Rr \times\Rr^+ \leqno(1.1.2)$$  be the moment map of the Hamiltonian $\Rr^2$-action
defined by the geodesic flow and by rotation.  The singular set of $P$ is the closed
conic set
$Z:= \{(r_o, \theta, 0, p_{\theta}): \theta \in [0, 2 \pi), p_{\theta} \in \Rr\}$,
i.e. $Z$ is the cone thru the equatorial geodesic (in either orientation).  The
image of $Z$ is the boundary of $B$; the map $P|_{T^*S_gS^2 - Z}$ is a trivial
$S^1 \times S^1$ bundle over the open convex cone $B_o$ (the interior of $B$).
 For each $b\in B_o$ , let
$H_1(F_b, \Z)$ denote the homology of the fiber $F_b := P^{-1} (b).$ This lattice
bundle is trivial since $B$ is contractible, so there exists a smoothly
varying homology basis $\{\gamma_1(b), \gamma_2(b)\} \in H_1(F_b, \Z)$ which equals
the unit cocircle $S^*_NS^2$ together with the fixed closed meridian $\gamma_M$ when
$b$ is on the center line $\Rr^+ \cdot (1,0)$. The action variables are given by
[CV.1,
\S 6]
$$I_1 (b) = \int_{\gamma_1(b)} \xi dx = p_{\theta},\;\;\;\;\;\;I_2 (b) =
\int_{\gamma_2(b)} \xi dx =\frac{1}{\pi} \int_{r_{-}(b)}^{r_{+}(b)} \sqrt{b_1^2 -
\frac{b_2^2}{a(r)^2}} dr + |b_2|\leqno(1.1.3)$$
where $r_{\pm}(b)$ are the extremal values of $r$ on the annulus $\pi(F_b)$ (with
$\pi : S^*_g S^2 \rightarrow S^2$ the standard projection). On the torus of meridians
in $S^*_g S^2$, the value of $I_2$ equals $\frac{L}{\pi}$ and it equals one on the
equatorial geodesic. So extended, $I_1, I_2$ are smooth homogeneous functions of
degree 1 on $T^*S^2$, and   generate
$2\pi$-periodic Hamilton flows.

It follows that the pair ${\cal I}:=(I_1, I_2)$ generates a global
Hamiltonian torus ($S^1 \times S^1$)-action commuting with the geodesic flow.
The singular set of ${\cal I}$ equals ${\cal Z}:= \{I_2 =\pm p_{\theta}\}$,
corresponding to the equatorial geodesics.
 The map
$${\cal I}: T^*S^2 - {\cal Z} \rightarrow \Gamma_o:= \{(x,y) \in \Rr \times \Rr^+ :
|x| < y\}\leqno(1.1.4)$$
is a trivial torus fibration.  Henceforth we often write $T_I$ for the torus
${\cal I}^{-1} (I)$ with $I \in \Gamma_o$ and let $\Gamma $
denote the closure of $\Gamma_o$ as a convex cone. 
 The symplectically dual angle variables
$(\phi_1 = \theta, \phi_2)$ then give, by definition,  the flow times (mod $2\pi$)
 along the orbits of $(I_1, I_2)$ from a fixed point on $F_b$, which we may take to be
the unique point lying above the intersection of the equator and the fixed meridian
on $F_b$ with the geodesic pointing into the northern hemisphere.  

So far, we have only assumed the metric to be $C^{\infty}$.  We now
 observe
 that if $g$ is real analytic, then so are $I_1, I_2.$  This is obvious in
the  case of $I_1$ and follows from the explicit formula (1.1.3) for $I_2.$
\qed

Since the metric norm function $|\xi|_g$ commutes with $I_1, I_2$,
it may be expressed as a function $H(I)$ of the action variables.
 Hamilton's equations for the geodesic flow then take the form
$$\dot{I}_k = 0, \;\;\;\;\;\;\;\dot{\phi}_{k} = \omega_k(I),\;\;\;\;\;(k=1,2) $$
where
$$\omega(I)=  \nabla_I H (I)\leqno(1.1.5)$$
is the {\it frequency vector} of the torus $T_I$ with action coordinates $I$.
The geodesic flow on  $T_I$ is then given by
$$G^t ( I, \phi) = (I, \phi + t \omega_I) \leqno(1.1.5a)$$
so that all the geodesics are quasi-periodic in action-angle coordinates.

The frequency vector $\omega_I$ is homogeneous of degree 0 on $T^*S^2 - 0$, and hence
is constant on  rays of torii $\Rr^+ T_I$. To break the $\Rr^+$ symmetry we restrict
to the level set
$\{ H(I) = 1\} \subset \Gamma_o$ in action space and view the frequency vector as the
map:
$$\omega : \{ H = 1\} \rightarrow \Rr^2.$$
Since $\nabla_I H (I)\;\;  \bot\;\; T_I (\{H=1\})$ the frequency map is more or less
the Gauss map of $\{H = 1\}$ (although it is not normalized to be of unit length).
As a map of the global action space, the frequency map is the Legendre transform
associated to $H$ (cf. [F.G, p.338]).
\bigskip

\noindent{\bf (1.1.6) Definition}~~~{\it We say that the simple surface of revolution
$(S^2,g)$ is {\em globally non-degenerate} if $\omega|_{\{H=1\}}$ is an embedding.}
\bigskip

This is the natural homogeneous analogue of the non-degeneracy condition of
[F.G, loc.cit] to the effect that the Legendre transform be a global diffeomorphism,
and has previously been studied in some detail by Bleher in the setting of
simple surfaces of revolution  [Bl,\S
6]. As will be seen in \S1.3, the curve $\{H = 1\}$ is the graph of a smooth function
of the form $I_2 = F (I_1)$ in the cone $\Gamma_o,$ and the non-degeneracy
condition (1.1.6) will follow as long as $F$ is a convex or concave function. In
fact, in the proof of the Theorem we will only need to use that $\{H = 1\}$ is
non-degenerate at the one point $(I_1, I_2) = (0,1).$  This is sufficient because
we assume the metric to be real analytic.

\begin{figure}[thb]\label{figure1}
\centering{\epsfig{file= 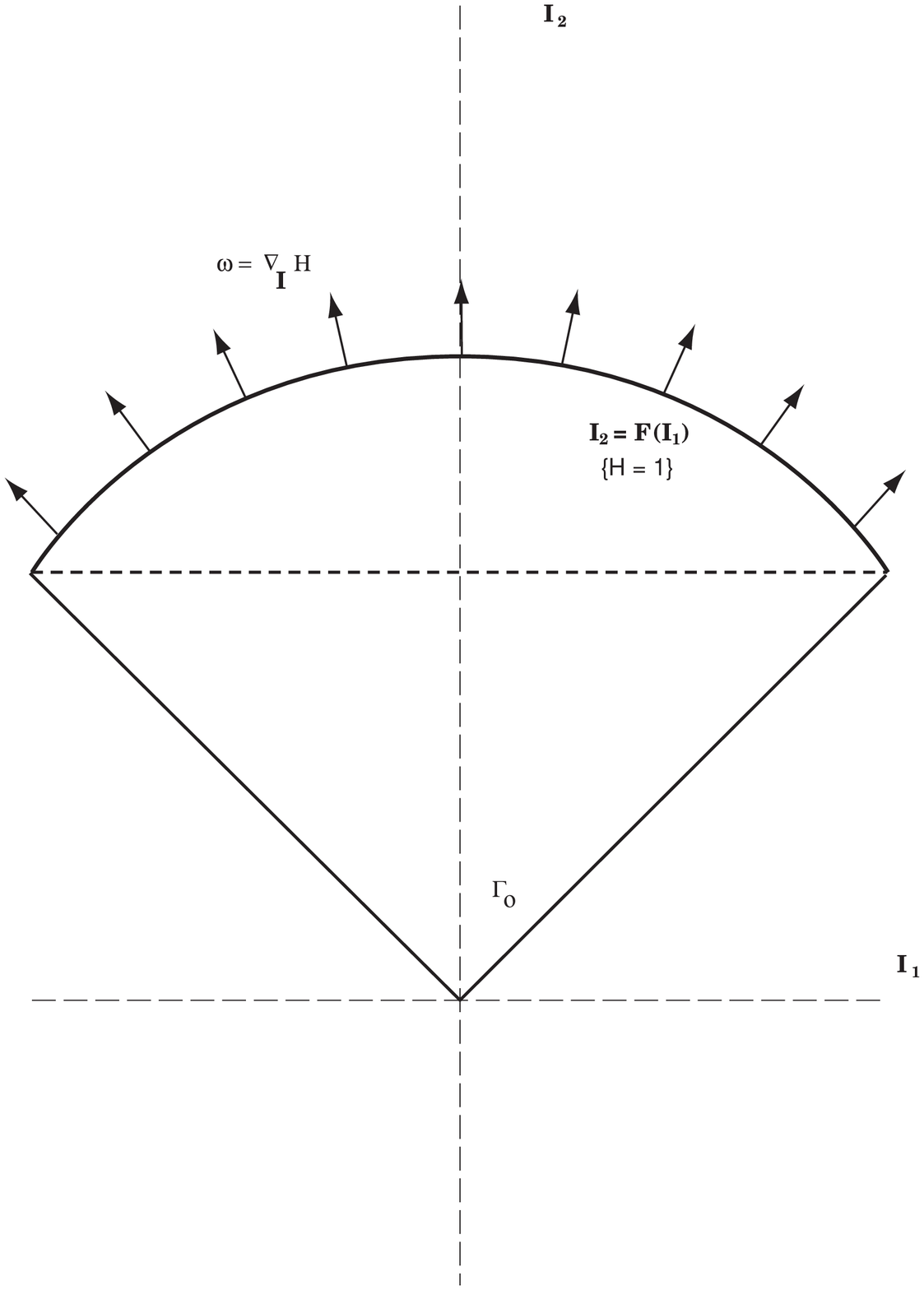, height= 3in}}
\caption{}
\end{figure}

\subsection{ Length spectrum and periodic torii }

We now come to the definition  of length spectrum
and simple length spectrum for a completely integrable geodesic flow.  We
first observe that  the orbit thru $(I, \phi)$ is periodic of period
$L$ if and only if
$$L \omega_I = M \in \Z^2 \leqno (1.2.1a)$$
for some $M \not= 0.$  The minimal positive such $L$ will be called the primitive
period; the corresponding $M$ is known as the  vector of winding numbers of the
torus $T_I$. $M$ parametrizes the homology class of  the closed orbit $\gamma$
since the latter has the form $\sum_{j=1}^2 M_j
\gamma_j(I)$ relative to the homology basis $\gamma_j(I)$.

Due to the homogeneity,
the period and vector of winding numbers are constant on the ray $\Rr T_I$.
By Euler's formula we then have
$$\nabla_I H \cdot I = \omega_I \cdot I = H$$
 hence the length is given in terms of the winding vector by 
$$L = \frac{M \cdot I}{H(I)}\leqno(1.2.1b)$$
or simply  $L= M \cdot I$ on the unit tangent bundle $H=1$. 

It is clear that the periodicity condition $L \omega = M$ is independent of
$\phi$. Hence,  all
of the geodesics on
$T_I$ are closed if any of them are.
( This also follows, of course, from the transitivity
of the torus action on each invariant torus.)  We therefore say:
\bigskip

\noindent{\bf (1.2.2) Definition}~~~{\it 

\noindent(a) A torus $T_I$ is
a {\em periodic torus} if all the geodesics on it are closed.\\
\noindent (b) The {\em period}
$L$ of the periodic torus is then the common period of its closed geodesics.\\
\noindent (c) The {\em length spectrum}
${\cal L}$ of the completely integrable system is the set of these
lengths.\\
\noindent (d) The completely integrable system has a  {\em simple length spectrum} if there
exist a unique periodic torus (up to time reversal) of each length $L \in {\cal L}$.}
\bigskip

In the last statement (d) we are referring to the canonical involution $\sigma: (x,
\xi) \rightarrow (x, - \xi)$, which reverses the orientation of the geodesics.  It
is obvious that if $T_I$ is a periodic torus of period $L$, then so is $\sigma
(T_I)$.  

Let $Per \subset S^*_gS^2-0$ denote the set of periodic points for the geodesic
flow on $S^*_gS^2$, i.e. the set of points which lie on a closed geodesic. It is a
union of periodic torii in $S^*_gS^2$ together with points along the equatorial
geodesics (which are degenerate torii).   The set of
all periodic points in $T^*S^2-0$ is then equal to $\Rr^+  Per.$  
Since the invariant torii are parametrized by the points $I \in \Gamma$ of action
space, it is convenient to parametrize $ Per$ 
by a subset of the level set $\{H(I) = 1\} \subset \Gamma.$
\bigskip

\noindent{\bf (1.2.3) Definition}{\it The set of points $I \in \{H=1\} \subset \Gamma$
such that $T_I \subset  Per$  will be called, with a slight abuse of notation,
the set of {\em  periodic} points on $\{H=1\}$ and will be denoted by ${\cal P}$. 
That is,
${\cal P} = \{ I \in \{H=1\}: \exists L \in \Rr^+, L \omega_I \in \Z^2\}.$}
\bigskip

The following proposition will be needed later on (Proposition (4.1.4).
\bigskip

\noindent{\bf (1.2.4) Proposition}~~~{\it If $(S^2,g)$ is non-degenerate (1.1.6),
then ${\cal P}$ is dense in $\{H=1\}$.}
\bigskip

\noindent{\bf Proof}: Let ${\cal Q}:= \omega ({\cal P})$ equal the image
 of ${\cal P}$ under the
frequency map $\omega$.  Then by definition, ${\cal Q}$ (for `rational points') 
is the projection to the curve $\omega(\{H=1\})$ of the integer
lattice in $\Z^2$. It is clear that ${\cal Q}$  is a dense set in
$\omega(\{H=1\})$; since $\omega$ is an embedding, ${\cal P}$ is dense in  $\{H=1\}.$
\qed
\bigskip

The next proposition  shows that our inverse result is valid for a residual set
of simple analytic surfaces of revolution.

\bigskip

\noindent{\bf (1.2.5) Proposition}~~~{\it Let ${\cal R}^* \subset {\cal R}$ be the
subset of metrics with simple length spectra.  Then ${\cal R}^*$ is a residual subset
of
${\cal R}$.}
\bigskip

\noindent{\bf Proof}: If $L, L' \in {\cal L}$, then there exist $M, M' \in
\Z^2$ and $I, I' \in {\cal Q}$ such that $L = M \cdot I, L' = M' \cdot I'.$
So $L=L'$ implies that $M \cdot I - M' \cdot I' = 0$ hence that the $I$-coordinates
of $I,I'$ are dependent over the rationals.  Since the length spectrum moves
continuously and non-trivially under deformations in ${\cal R}$, such a dependence
for fixed $M,M'$ can only hold on a closed nowhere dense set.  The proposition
follows.\qed

\subsection{First return times and angles}

 Let us  consider more carefully the
geometric interpretation of the Clairaut integral on a torus $T_I \subset S^*_g
S^2.$  Since $I_1$ and $H= |\xi|$ are independent commuting coordinates, and since
there are global action-angle variables, the different invariant torii in $S^*_g S^2$
are parametrized by the values  $I_1 = \iota \in [-1,1]$ of the Clairaut integral 
along
$\{H=1\}$. Indeed by (1.1.3), the second action coordinate $I_2$ is determined from
$I_1$ on 
$\{H = 1\}$ by the formula 
$$I_2 = F(I_1):= |I_1| + \frac{1}{\pi} \int_{ r_{-}(I_1)}^{r_{+}(I_1)}
 \sqrt{1 - \frac{I_1^2}{a(r)^2}} dr \leqno(1.3.1) $$
where $r_{\pm}(I_1)$ are the roots of $I_1^2 = a(r)^2.$  
 The torus in $S^*_g S^2$ with $I_1 = \iota$ is therefore $T_{(\iota, F(\iota))}$,
which we will denote simply by $T_{\iota}.$ 
The
projection of $T_{\iota}$ to $S^2$ is an annulus of the form $r_+(\iota) < r <
r_{-}(\iota)$. the geodesics on $T_{I}$ project to $S^2$
as almost periodic curves oscillating between the two extremal parallels $r=
r_+(\iota)$ and
$r= r_{-}(\iota).$ 

 We observe then that
$T_{\iota}$ contains a unique geodesic $\gamma_{\iota}$ which passes thru the
intersection of the reference geodesics $\gamma_M$ and $\gamma_E$;  $\iota$
 equals $a(r_o) cos \alpha$ where $\alpha (\iota)$ is the
angle between $\gamma_{\iota}$ and $\gamma_M$.  In other words, $\alpha(\iota)$ is the
common angle with which the geodesics on $T_{\iota}$ intersect the meridians as they
pass thru the equator in the direction of the northern hemisphere.
Since the length $L_E$ of the equator $\gamma_E$ equals $2 \pi \sqrt{a(r_o)}$, and
since this length is  a symplectic invariant of the geodesic flow, the coordinates
$\iota$ and $\alpha$ are related in an essentially universal fashion.  Hence,  
either $I_1$ or  $\alpha \in (0,  \pi)$ could be used as an  
action coordinate on $\{H=1\}$; $\alpha$ is perhaps more geometric but $I_1$ is
more convenient in calculations. 

 The picture is the same for
  any invariant torus $T_I$:  Under the $\Rr^+$ action on $T^* S^2-0$,
it scales to a torus $T_{\iota} \subset S^* S^2$ and  all features
of its geometry are identical to that of $T_{\iota}$.  Thus it carries
a unique (parametrized) geodesic $\gamma_I$ such that $\gamma_I(0)$ is at
the intersection $\gamma_E \cap \gamma_M$,
and so that $\gamma_I'(0)$ points to the northern hemisphere. 
The initial
angle variables of $(\gamma_I(0), \gamma_I'(0))$ are therefore $(\phi_1(0), \phi_2(0))
=(0,0).$ At time $t$ we denote the angle variables of $(\gamma_I(t), \gamma_I'(t))$ by
$(\phi_1(t), \phi_2(t))$ where $\phi_1$ essentially measures the meridianal angle
and $\phi_2$ measures the equatorial angle.
We then introduce the following `first return times':
\bigskip

\noindent{\bf (1.3.2) Definition}{\it \\ 
\noindent(Ei) The equatorial first return time is the minimal time $\tau_E(I)> 0$ such
that $\phi_1 (\tau_E (I)) = 2 \pi$;\\
\noindent(Eii) The equatorial first return angle  $\omega_E(I)
:= \phi_2 (\tau_E(I))$  is the  change in angle along the equator of a geodesic
on $T_I$ leaving $\gamma_E \cap \gamma_M$ at $t=0$, upon its first return time to
$\gamma_E$;\\
\noindent (Mi) The meridianal first return time is the minimal time $\tau_M(I)$
such that $\phi_2 (\tau_M (I)) = 2 \pi.$\\
\noindent(Mii) The meridianal first return angle is the angle change $\omega_M(I):=
\phi_1 (\tau_M(I))$ along the meridian of a geodesic on $\gamma_M$ leaving
$\gamma_E \cap \gamma_M$  at $t=0$,  upon its first return time to $\gamma_M$.}
\bigskip

  The terminology `first return time' is taken from dynamics. 
Note that $\phi_1 \equiv 0 (mod 2\pi)$ is the equation of the curve on $T_I$ which
lies over the equator.  Hence, $\tau_E(I)$ is the time of first return of
$\gamma_I(t)$ to the equator in the direction of the northern hemisphere.
This is
actually the second time of intersection of $\gamma_I$ with the equator, the first
one occurring when
$\gamma_I(T)$ is heading to the southern hemisphere.  This intersection is not
in the projection of $\phi_1 \equiv 0 (mod 2\pi)$.
Similarly, $\phi_2 \equiv 0 (mod 2\pi)$ is the equation of the fixed meridian
$\gamma_M$,  and so $\tau_M$
is the time of first return to the arc $\gamma_M$ (half of the closed geodesic).

\begin{figure}[thb]\label{figure1}
\centering{\epsfig{file= 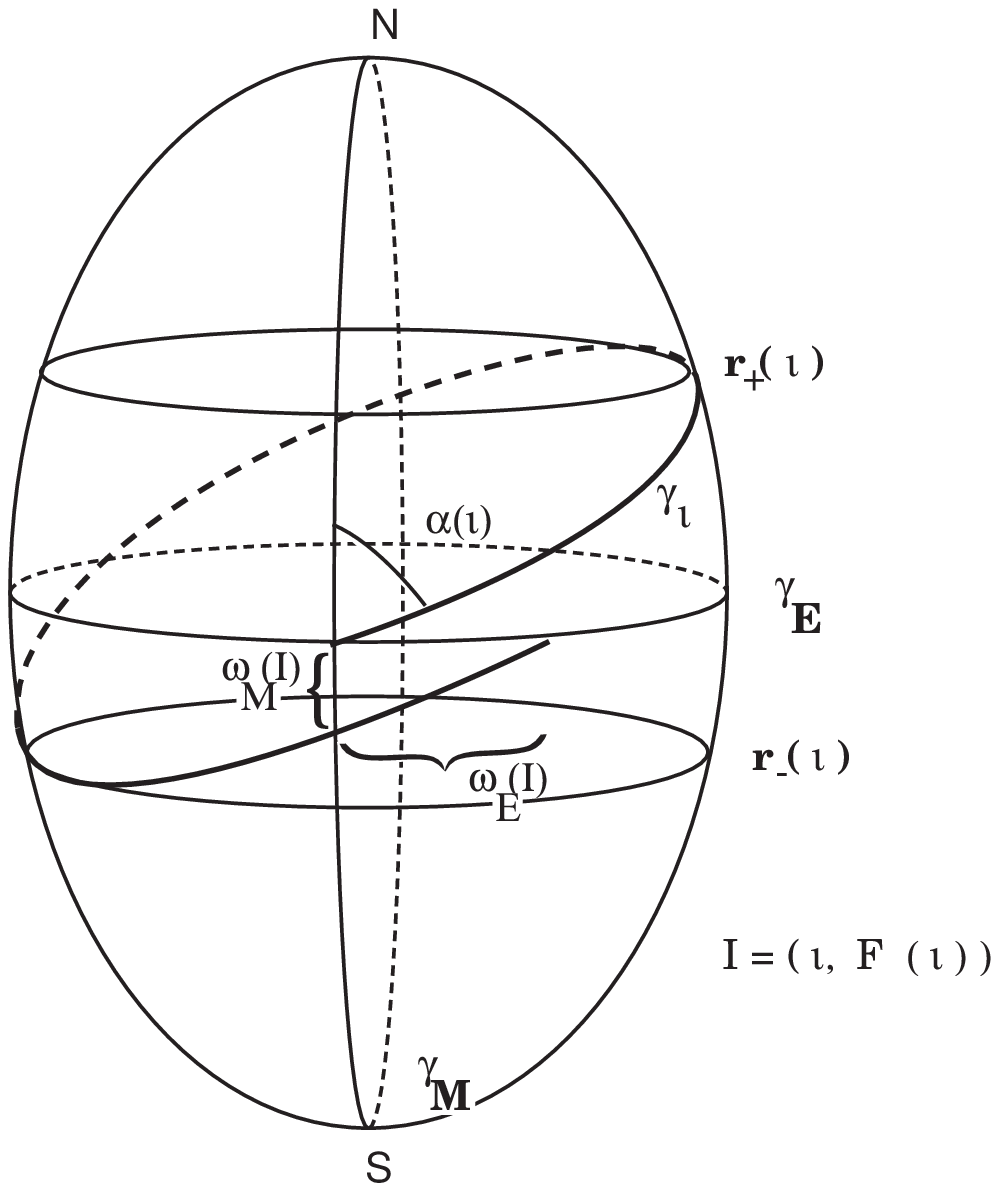, height= 3in}}
\caption{}
\end{figure}

The following gives some relations between the various angles and return times.
\bigskip

\noindent{\bf (1.3.3) Proposition}~~~{\it Let $\omega_I = (\omega_1, \omega_2)$
be the frequency vector of the invariant torus $T_I$. Then:\\
\noindent(a) $\tau_E\;\; \omega_2 = \omega_E$;\\
\noindent(b) $\tau_M \;\; \omega_1 = \omega_M$;\\
\noindent(c)  
$2\pi \frac{\omega_1(I)}{\omega_2(I)} = \omega_M,
\;\;\;\;\;\;2\pi \frac{\omega_2(I)} {\omega_1(I)}  
=\omega_E.$ }
\bigskip

\noindent{\bf Proof}: The equation of the geodesic $(\gamma_I(t), \gamma_I'(t))$
on $T_I$ beginning at $(\phi_1, \phi_2)$ in angle variables is  $(\phi_1 + t
\omega_1(I), \phi_2 +t \omega_2(I))$.  When the torus is projected to the base,
then $\phi_1$ measures the meridianal angle and $\phi_2$ measures the equatorial
angle.  So by definition of return times,
\medskip

\begin{tabular}{l} $(\phi_1 + \tau_E \omega_1, \phi_2 + \tau_E \omega_2) = 
(\phi_1 + 2 \pi, \phi_2 + \omega_E) $\\
$(\phi_1 + \tau_M \omega_1, \phi_2 + \tau_M \omega_2) = 
(\phi_1 +  \omega_M ,\phi_2 + 2 \pi). $
\end{tabular} \medskip

The statements in the proposition follow immediately.
\qed 
\bigskip

These first return times (and angles) are closely related to the (non-linear) Poincare
maps of the geodesic flow.  
 We recall that for each closed geodesic $\gamma$, the
Poincare map ${\cal P}_{\gamma}$ is defined as the first return map of the geodesic
flow, restricted to a symplectic transversal $S_{\gamma} \subset S^*_g S^2$ (surface
of section).  It is well-known ${\cal P}_{\gamma}: S_{\gamma} \rightarrow S_{\gamma}$
is a symplectic map [K].  Since the symplectic form on a cotangent bundle equals
$d \alpha$ (with $\alpha$ the action form), it follows that ${\cal P}_{\gamma}^*
(\alpha) - \alpha$ is a closed 1-form on $S_{\gamma}.$ Since $S_{\gamma}$ may be
assumed contractible, it follows that there exists a function $\tau_{\gamma}$ such
that  $[{\cal P}_{\gamma}^* (\alpha) - \alpha]|_{S_{\gamma}} = d \tau_{\gamma}.$
Since the integral of $\alpha$ over an arc of a unit speed geodesic just gives its
length, $\tau_{\gamma}$ is the first return time of geodesics near $\gamma$ to
$S_{\gamma}.$

In particular, let $\gamma = \gamma_E$ be the equator (in one of its orientations),
and let $S_E$ denote a symplectic transversal at the point $(\gamma_E(0),
\gamma_E'(0))$ thru the fixed meridian $\gamma_M$.  Since $\gamma_M$ is transverse
to the equator, we may define $S_E$ to consist of a small variation of $\gamma_E'(0)$
moved up and down a small arc of $\gamma_M$.  We see then that the first return
time $\tau_{\gamma_E}$ is precisely the first return time $\tau_M$ defined above.
We also observe that the foliation of $S^*S^2$ by invariant torii restricts to
a foliation of $S_E$ by invariant circles for ${\cal P}_{\gamma_E}$, closing in
on the center point where $\gamma_E$ intersects $S_E$.  As noted above, the
action coordinate $I_1 = \iota$ gives a natural action (radial) coordinate on
$S_{\gamma}$.  Since $S_{\gamma} \subset S^*_g S^2$, the $\Rr^+$ homogeneity is broken
and we may reformulate the  twist condition (0.1 (iii)) as
follows:
\bigskip

\noindent{\bf (1.3.4) Proposition}~~~{\it The Poincare map  ${\cal P}_{\gamma_E}$
is a twist map of $S_E$ iff $\omega_M ' \not= 0$, where $\omega_M' =
\frac{\partial}{\partial \iota} \omega_M.$}
\bigskip

\noindent{\bf Proof}:  
  The coordinates $(I_1 = \iota, \phi_1)$ restrict to a system of
symplectic coordinates on $S_E$, in terms of which the Poincare map takes the form
$${\cal P}_{\gamma_E} (\iota, \phi_1) = (\iota, \phi_1 + \tau_M \omega_1) =
(\iota, \phi_1 + \omega_M).$$
By definition it is a twist map if $\omega_M'\not= 0$  in
a neighborhood of $\iota = 0$ in $S_E$. \qed 
\bigskip

For background on the twist condition in a related context see  [F.G][P].

The situation for the other periodic orbits is different since they come in
one-parameter families.  Thus, for the (closed) meridian geodesic $\gamma_M$,
a transversal $S_M$ is given by the equator $\gamma_E$ and a small variation of 
$\gamma_M'(0)$ along it. The foliation by invariant torii restricts to $S_M$
to a foliation by invariant lines (non-closed curves), including the curve of
closed geodesics thru $\gamma_M$. The Poincare map
$P_{\gamma_M}$ is then of parabolic type; the first return time is $\tau_E$ above.

For the inverse problem it will be necessary to have expressions for these
return times and angles in terms of the metric. This will also make the twist
condition more transparent.
 For ease of quotation, it is convenient to make a change of dependent and
independent variables, following [Besse] and Darboux [D]. Equivalent expressions
in the original polar coordinates can also be easily derived, and will be given
below.
\bigskip

\noindent{\bf (1.3.5) Proposition}~~~{\it Suppose that $(S^2,g)$ is a simple surface
of revolution.  Then there exists a coordinate system $(u, \theta)$ on $U$, with
$sin u = a(r) $ and a smooth function $f$ on $[-1,1]$ such that
$f(1) = 1, f(-1) = 1$ and such that
$$g = [f(cos u)]^2 du^2 + sin^2 u d\theta^2.$$}
\medskip

\noindent{\bf Proof}: First, define
$$b(r) = \left \{ \begin{array}{ll} sin^{-1} (a (r )) & r \in [0, r_o] \\
\pi - sin^{-1} (a (r )) & r \in [r_o, L] \end{array} \right .$$
and 
$$c(v) = \left \{ \begin{array}{ll}  (a |_{[0, r_o]})^{-1} (\sqrt{1 - v^2} ) & v \in
[0, 1] \\
(a |_{[r_o, L]})^{-1} (\sqrt{1 - v^2} ) & v \in [-1, 0] \end{array} \right.$$
Then define $f : (-1, 1) \rightarrow \Rr$ by
$$ \left\{ \begin{array}{l} f(v) = \frac{v}{a'[c(v)]},\;\;\;(v\not= 0) \\
f(0) = \frac{1}{- a^{''}(r_o)} \end{array} \right .$$

Since $b(r) =u , c(cos u) = b^{-1}(u) = r$ we have $a(r) = sin u, 
a'[ c(cos u)] dr = cos u du$ and $b'(r) dr = du.$  The smoothness properties of
$f$ follow from those of $a$ and from the fact that $a(r)$ and $sin u$ have
the same qualitative shapes [Besse, loc.cit.]. \qed
\bigskip

The  geometric result is the following [Besse, Theorem 4.11]:
\medskip

\noindent{\bf (1.3.6) Proposition}~~~{\it In the above coordinates, the equator
$\gamma_{E}(s) = (u(s), \theta(s))$  has
the equation
$$u(s) \equiv \frac{\pi}{2},\;\;\;\;\; \theta(s) = s.$$
Any other geodesic $\gamma(s) = (u(s), \theta(s))$ in $U$ is contained between
two parallels $u = i$ and $u = \pi - i$ and the angle $\theta(i)$ between
two consecutive points of contact with these parallels is given by:
$$\theta(i) = sin i \int_i^{\pi - i} \frac{f(cos u)}{sin u (sin^2 u - sin^2
i)^{\half}} du.$$ The length of this arc of $\gamma$ is given by
$$s(i) =  \int_i^{\pi - i} \frac{ sin(u) f(cos u)}{ (sin^2 u - sin^2
i)^{\half}} du.$$}
\medskip

\begin{figure}[thb]\label{figure3}
\centering{\epsfig{file= 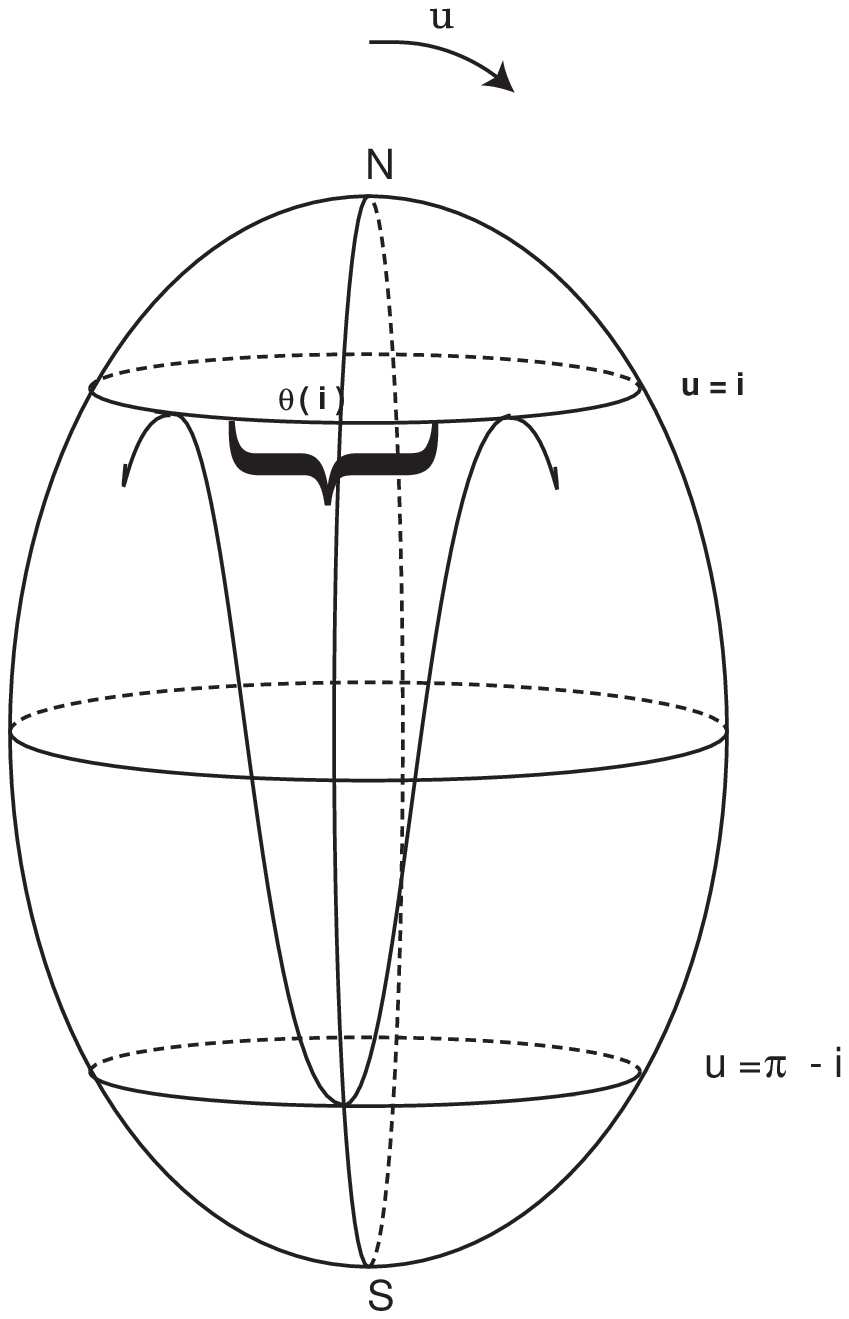, height= 3in}}
\caption{}
\end{figure}

\newpage

\noindent{\bf Sketch of Proof:} Using the Clairaut integral, the equations of the
geodesic have the form:
$$\left \{ \begin{array}{l} \frac{d\theta}{ds} = \frac{sin i}{sin^2 u} \\
  \frac{d\theta}{du} = \frac{sin (i) f(cos u)}{sin u (sin^2 u - sin^2 i)^{\half}}\\
\frac{ds}{du}=\frac{sin (u) f(cos u)}{ (sin^2 u - sin^2 i)^{\half}} \end{array}
\right .$$
The formulae above follow by integration. \qed

The following is geometrically obvious:
\bigskip

\noindent{\bf (1.3.7) Proposition}~~~{\it Let $T_I \subset S^*_gS^2$ denote an
invariant torus with $H(I) = 1$ and let $i(I)$ be the $u$-coordinate of the 
extremal parallel closest to $N$ in the projection of $T_I$ to $S^2$.  Then:\\ 
\noindent(i) $\tau_E(I) = 2 s (i(I))$;\\

\noindent (ii) $\omega_E(I) = 2 (\theta (i(I)) - 1)$;\\

\noindent (iii) $\tau_E(I) = \frac{1}{\pi} \int_{ r_{-}(I_1)}^{r_+(I_1)}
 (1- \frac{I_1^2}{a(r)^2})^{-\half} dr$\\

\noindent (iv) $ \omega_E(I) =  - 1 + \frac{I_1}{\pi} \int_{
r_{-}(I_1)}^{r_+(I_1)} a(r)^{-2}
 (1 - \frac{I_1^2}{a(r)^2})^{-\half} dr .$ }
\bigskip

\noindent{\bf Proof:}

\noindent(i) By definition, $\omega_E(I)$ is the change in angle along
the equator between a geodesic $\gamma_I$  on $T_I$, starting at the equator on a
fixed meridian and heading towards the northern hemisphere, and the
fixed meridian, upon second intersection with the equator.  We can view this arc of
the geodesic as consisting of three pieces: one from the equator to the
northern extremal parallel, one on the `back-side' between the two extremal parallels,
and one on the `front-side' from the southern parallel to the equator. Since the
lengths of the `front-side' arcs are unchanged by rotation, we can rotate one until
the two make up a smooth geodesic arc between the parallels.  The length of the
geodesic arc is therefore twice that of an arc between the parallels, i.e.
$\tau_E(I) = 2 s (i(I)).$  

\noindent(ii) For the angle change:  In the same way, the change in $\theta$ along
this arc of
$\gamma_I$ is the change in $\theta$ of two arcs between the extremal parallels.
We subtract $1$ since $\omega_I$ measures the addition to one full revolution.

\noindent(iii)  Since $sin (u) = a (r)$ and $f(cos u) du = dr$, we have 
$$2 s(i(I)) = \pi I_1 \int_{i(I)}^{\pi - i(I)} \frac{dr}{\sqrt{ 1 -
\frac{I_1^2}{a(r)^2}}}.$$ 

\noindent(iv)  From Propositions 1.3.2 we have that $\frac{\omega_1}{\omega_2}
= \frac{1}{2\pi} \omega_E.$  Since
$H(I_1, I_2) = 1$ implies that $\omega_1 + F'(I_1) \omega_2 = 0,$ we
get from (1.3.1) that $\omega_E = - 2 \pi F'(I_1).$ \qed
\bigskip

\noindent{\bf (1.3.8) Corollary}~~~{\it The non-degeneracy condition (1.1.6) is
satisfied if ${\cal P}_{\gamma_E}$ is globally twisted, i.e. if $\omega_M ' > 0$
or $\omega_M' < 0.$.}
\bigskip

\noindent{\bf Proof}: In these cases, $F$ is convex (or concave).  Since the
set $\{H = 1\}$ is the graph of $F$ in $\Gamma_o$, 
 the Gauss map (and hence the frequency map) 
 is an embedding.\qed
\bigskip

\noindent{\bf Remark} Both cases, of concavity and convexity, occur for
ellipsoids of revolution, see [Bl].  The separating case of the round sphere
is of course degenerate.

\subsection{ Classical  Birkhoff invariants }

The classical Birkhoff normal form of  Hamiltonian $H$ near a non-degenerate
periodic orbit $\gamma$ is a germ of a completely integrable system to which $H$ is
symplectically equivalent in a `formal neighborhood' of $\gamma$ (see e.g.
 [F.G] for a detailed discussion).    In the case at hand,
where $H$ is already completely integrable, the Birkhoff normal form is simply
$H$ itself expressed in terms of action-angle variables.
The `Birkhoff invariants' of the Hamiltonian at a torus $T_{I_o}$ may
then be identified with the germ of  $H(I_1, I_2)$ or of $\omega_I$ at $I= I_o.$  

Since $H$ is homogeneous of degree 1, it is equivalent and somewhat clearer
to define the Birkhoff invariants after first breaking the $\Rr^+$ symmetry. That is,
we would like to introduce a `base' to the cone $\Gamma_o$. The most natural one
may appear to be the energy level $\{H=1\}$; but for the purpose of calculating wave
invariants at $T_{I_o}$ it is more convenient to use
the tangent line
$\omega_{I_o} \cdot (I - I_o) = 0$ at a point $I_o \in \{H=1\}.$ 

Let us first consider the level set $\{H=1\}$ as the transversal. The homogeneous
function
$H$ is obviously determined by the curve  $H(I_1, I_2)=1$ whose equation is given by
(1.3.1).
Hence we can define the Birkhoff invariants at a  torus $T_{I_o}$ to be the
Taylor coefficients of the function $F'$ at $I_o \in \{H=1\}$.  By (1.3.7 (iv))
it is equivalent to put:
\bigskip

\noindent{\bf (1.4.1) Definition}~~~{\it The (first) Birkhoff invariants of $H$ at an
invariant torus $T_{\iota}$ with  $\iota \in \{H = 1\}$ are the Taylor coefficients
of   $\omega_E$ at $\iota$ in the coordinate $I_1$.}
\bigskip

Secondly, let us consider the  tangent lines as transversals:
 We fix a point $I^o \in \{H=1\}$, let 
 $\omega^o$ denote the common frequency vector of the ray of torii $\Rr^+ T_{I^o}$ and
put
$$I \cdot \omega^o := \sum_{k=1}^2 \omega_k^o I_k.\leqno(1.4.2) $$
The equation of  
the tangent line to $\{H = 1\}$ at $I^o$ in the action cone $\Gamma_o$ 
is then given by  $\omega^o \cdot I = 1$. 
Note that $I \cdot \omega^o$ is homogeneous of degree 1 and hence equals $H(I)$ along
the ray $\Rr I^o$; consequently  it is elliptic (non-vanishing) in a conic
neighborhood $W_o\subset \Gamma_o$ of it. The conic neighborhood will be
parametrized in the following way: we fix a basis (i.e. a non-zero vector)  $v$ of
the line $I \cdot \omega^o  =0$, and define the map
$$(\rho, \xi) \rightarrow \rho (I^o + \xi v), \;\;\;\;\;\;\;\xi \in (-\epsilon,
\epsilon).\leqno(1.4.3a)$$
For sufficiently small $\epsilon$, this map sweeps out a conic neighborhood $W_o$ of
$I^o$ with inverse given by 
$$\rho = \omega^o \cdot I, \;\;\; \xi v_j:= \frac{I_j}{I \cdot \omega^o} -
I_j^o.\leqno(1.4.3b)$$ 
Since $ H (I_1, I_2) = (\omega^o \cdot I) \;\;H(\frac{I_1}{\omega^o \cdot I},
 \frac{I_2}{\omega^o \cdot I})$ and since $(\frac{I_1}{\omega^o \cdot I},
 \frac{I_2}{\omega^o \cdot I}) \in \{\omega^o \cdot I = 1\}$ we may write
$$ H (I_1, I_2) = \rho h_{I^o} (\xi)\leqno(1.4.4a)$$
where $h_{I^o}$ is the function on $W_o \cap \{\omega^o \cdot I = 1\}$ defined by
$$h_{I^o}(\xi):= H(I^o + \xi v).
\leqno(1.4.4b)$$
The $C^{\infty}$ Taylor expansion of $h_{I^o}(\xi)$ around $\xi = 0$ is then a
symplectic invariant of $H$.
\bigskip

\noindent{\bf (1.4.5) Definition}~~~{\it The second (tangential) classical Birkhoff
invariants of
$H$ associated to the periodic torus $T_{I^o}$ are the Taylor coefficients
$h_{I^o}^{\alpha}(0).$}
\bigskip

In the real analytic case the Taylor coefficients determine $h_{I^o}$ and hence
$H$ by homogeneity.  It is more or less obvious that the first and second
Birkhoff invariants also carry the same information in the $C^{\infty}$ case.
To be sure, we prove:

\bigskip

\noindent{\bf (1.4.6) Proposition}~~~{\it The first Birkhoff invariants canonically
determine  the tangential Birkhoff invariants and vice versa.}
\bigskip

\noindent{\bf Proof}: 

By definition
$$h_{I^o}(\xi) = H(I^o + \xi v).$$ Therefore,
$$  H(h_{I^o}(\xi)^{-1} (I^o + \xi v)) = 1$$
or equivalently 
$$h_{I^o}(\xi)^{-1}(I^o_2  + \xi v_2 )= F(h_{I^o}(\xi)^{-1}(\xi v_1 + I_1^o)).$$
Writing  $u = \frac{I^o_1 + \xi v_1}{h_{I^o}}$,
this says
$$\frac{F(u)}{u} = - \frac{I^o_2 + \xi v_2}{I^o_1 + \xi v_1} .$$ 
Hence the knowledge of Taylor coefficients of $F$ is equivalent to knowledge of
the Taylor coefficients of $h_{I^o}$.\qed
\bigskip

\begin{figure}[thb]\label{figure4}
\centering{\epsfig{file= 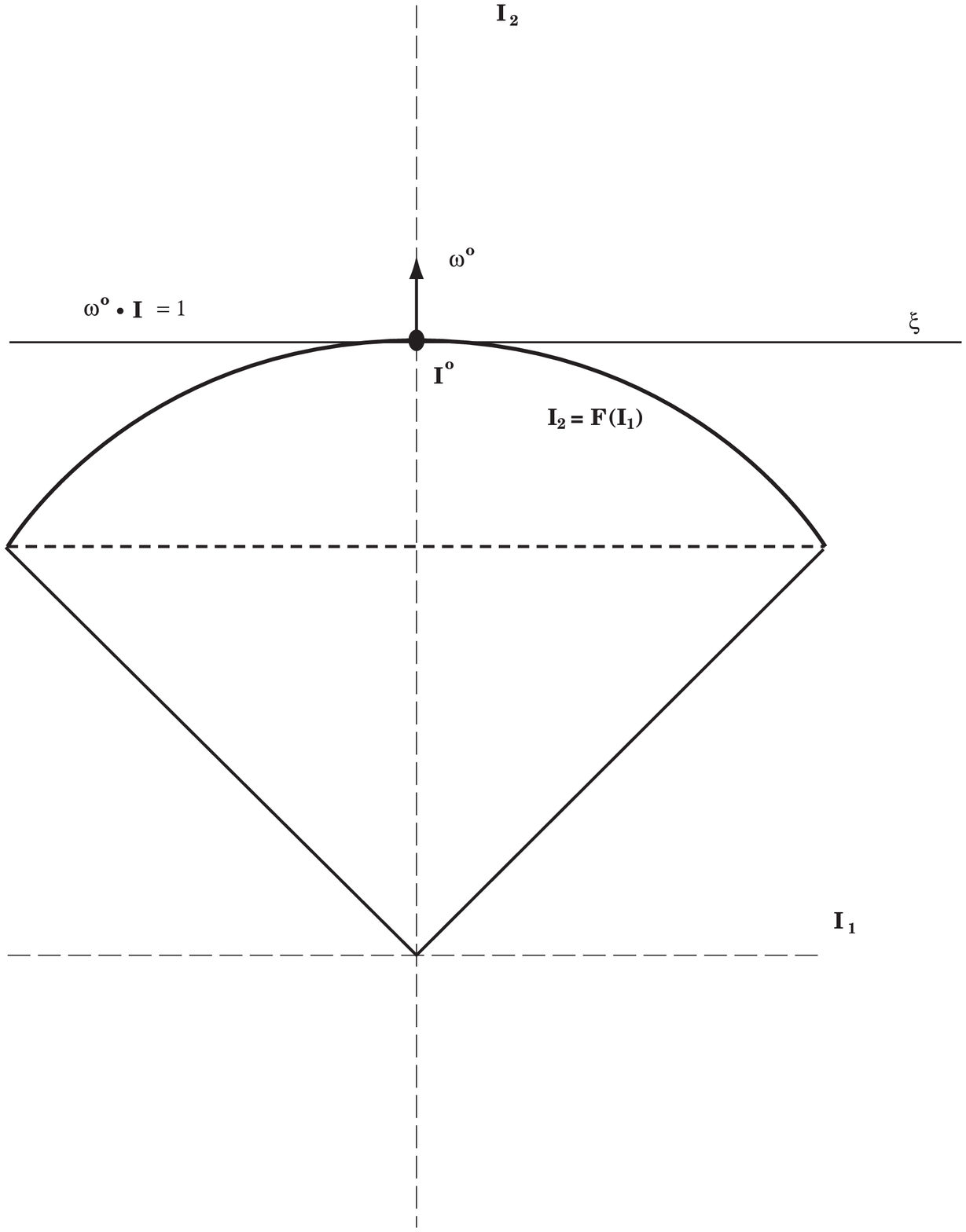, height= 3in}}
\caption{}
\end{figure}

We may reformulate the non-degeneracy condition (0.1 (iii)) in terms of $h_{I^o}$
where $T_{I^o}$ is the meridian torus:
\bigskip

\noindent{\bf (1.4.7) Proposition}~~~{\it $(S^2,g)$ satisfies the non-degeneracy
condition (0.1 (iii)) as long as $\alpha:= h_{I^o}''(0) \not= 0.$}
\bigskip

\noindent{\bf Proof}: By definition, $h_{I^o}(\xi) = H(I^o + \xi v)$. At the
meridian torus, $v = (1,0)$ so $h_{I^o}(\xi) = H(I^o + (\xi, 0))$ and 
$h_{I^o}'(\xi) =\frac{\partial}{\partial I_1} H (I^o + (\xi, 0))  = \omega_1(I^o +
(\xi, 0)$.  Hence
$$h_{I^o}''(0) =  \frac{\partial}{\partial I_1} \omega_1 (I^o).$$
Since $\omega_1(I^o) = 0$ it follows that 
$$\frac{\partial}{\partial I_1} \omega_M (I^o) =
\frac{\omega_1'(I^o)}{\omega_2(I^o)}.$$ Hence, the condition  $h_{I^o}''(0) \not= 0$
is equivalent to the condition that $\frac{\partial}{\partial \iota} \omega_M (I^o)
\not= 0.$  The equivalence of this to (0.1 (iii)) is proved in Propositions (1.3.4)
and again in Corollary (1.3.8).
\qed

\subsection{Symplectic conjugacy of geodesic flows}

The Birkhoff normal form of a Hamiltonian $H$ at a closed orbit (or family of closed
orbits) is a symplectic conjugacy invariant of $H$ in a neighborhood of the
orbit(s).  Hence the global Birkhoff normal form $H(I_1, I_2)$ of a completely
integrable Hamiltonian is a symplectic conjugacy invariant.  The purpose of this
section is to show that it is a complete conjugacy invariant.  We begin by showing
that the homogeneous Hamiltonian torus actions commuting with geodesic flows
of simple surfaces of revolution are all symplectically equivalent: 
\bigskip

\noindent{\bf (1.5.1) Proposition}~~~{\it Suppose $(S^2,g_1)$ and $(S^2,g_2)$ are
smooth surfaces of revolution of simple type and let $(I_1, I_2)$ resp. $(J_1, J_2)$
denote their global action variables as above.  Then there exists a homogeneous
symplectic diffeomorphism  $\chi : T^*S^2 \rightarrow T^*S^2$ such that $\chi^*J_i =
I_i$.}
\bigskip

\noindent{\bf Sketch of Proof :} Let $\phi_i$, resp. $\psi_i$ be the dual
angle variables on
$(S^2,g_1)$ resp. $(S^2,g_2)$.  Except on the equators of $(S^2, g_i)$ a point in
$T^*S^2$ is uniquely specified by its action-angle coordinates.  Define $\chi$ to
be the identity map in action-angle coordinates with respect to the two metrics.
It is obvious that $\chi$ is a homogeneous symplectic diffeomorphism on the
complement of the equators, so to prove the Corollary it suffices to show that
$\chi$ extends to the equators with this property. 

Since $\chi$ is homogeneous of degree 1, it is necessary and sufficient to define 
it on the unit cotangent bundles.  Moreover, since it commutes with the Hamilton flow
of
$p_{\theta}$ on the regular set, its extension must also do so. 
Hence it must be the lift of a  diffeomorphism $\bar \chi$ on the orbit space ${\cal
O}: =S^*S^2/S^1$ of the rotation action.  This action is free so the natural
projection $p : S^*_g S^2 \rightarrow {\cal O}$ must be diffeomorphic to the standard
projection from $S0(3) \rightarrow S^2.$ The image of the torus foliation defined
by level sets of $p_{\theta}$ is a singular foliation of ${\cal O}$ formed by 
level sets of the function $\bar p_{\theta}$ induced by $p_{\theta}$ on ${\cal O}$,
and the two singular points $o^{\pm}$ are the images of the equators $\gamma_E^{\pm}$.
Since $\bar p_{\theta}$ is, by assumption, a perfect Morse function on ${\cal O} \sim
S^2$ for each metric, the quotient map $\bar \chi$ on the punctured quotient ${\cal
O}-\{o^{\pm}\}$ of $\bar \chi$ extends smoothly to the completion.   It follows
that $\chi$ extends smoothly as a rotationally equivariant map on the completion
of $S^*S^2 - \{ \gamma_E^{\pm}\}$ and the homogeneous extension must be symplectic.
(See [CV.1] for more on the behaviour near the poles).
\qed
\bigskip

\noindent{\bf Remark}: In fact, this proposition can be sharpened to say: there exists
only one  homogeneous Hamiltonian $S^1 \times S^1$ action on $T^* S^2-0$ (up to
symplectic equivalence).  This follows from a homogeneous analogue of the Delzant
classification of completely integrable torus action on compact Kahler manifolds.
We hope to report on more general results of this kind at a later time.

Thus, the torus actions defined above on the cotangent bundles of simple surfaces of
revolution are always symplectically equivalent. The question arises when the
geodesic flows are symplectically equivalent.  The answer can be given by expressing
the norm functions $|\xi|_g$ of the metrics in terms of the global action variables.
Before doing so, we note that the action variables are not quite uniquely defined
above because the choice of generators $\gamma_i(b)$ is not unique.  For instance,
one might have permuted the roles of $N$ and $S$.  Hence we have:
\bigskip

\noindent{\bf (1.5.2) Proposition}~~~{\it Let $(S^2,g_i)$ be a simple surfaces of
revolution, and let $(I_1,I_2)$ resp. $(J_1, J_2)$ be the global action variables
defined above.  Let $|\xi|_{g_1} = H_1(I_1, I_2)$ resp. $|\xi|_{g_2} = H_2(J_1, J_2)$
be the expressions of the metric norms of $g_i$ in terms of action variables.  Then:
the geodesic flows of $(S^2,g_i)$ are homogeneously symplectically equivalent if and
only if there a linear map $A = \left( \begin{array}{ll} a_{11} & a_{12} \\
a_{21} & a_{22} \end{array} \right ) \in SL(2, \Z)$ such that $H_1 = H_2 \circ A.$}
\bigskip

\noindent{\bf Proof} If such a choice exists, then the map $\chi$ above obviously
defines a symplectic conjugacy.

Conversely, suppose the geodesic flows are symplectically conjugate by a
homogeneous symplectic diffeomorphism $\chi$, i.e. $\chi^* H_2(J_1, J_2)
= H_1(I_1, I_2).$  Then $\chi^* J_i$ are global
action variables for the geodesic flow of $(S^2, g_1)$.  But global action
variables are almost unique: they correspond to a trivialization of the lattice
bundle $H_1(F_b, \Z)$ [D][G.S]. Therefore there exists 
 $A \in SL(2,\Z)$ so that  $\chi^*J = A \cdot I.$ Then $H_1(I) = H_2(A \cdot
I)$. \qed
\bigskip

The following gives a  geometric interpretation of the conjugacy condition:
\bigskip

\noindent{\bf (1.5.3) Proposition}~~~{\it Suppose $g_1, g_2 \in {\cal R}$.  Then their
geodesic flows are symplectically equivalent if and only if their equatorial first
return times $\tau_E$ and angles $\omega_E$ are equal.}
\bigskip

\noindent{\bf Proof}:  Suppose first that the flows are symplectically equivalent.
 After choosing compatible bases for the homology we may then
assume that the expressions for $H_1$ and $H_2$ in global action-angle variables
are the same.  Then the frequency maps are the same and hence by Proposition
(1.3.3) the equatorial return angles are the same. Also, the Poincare maps 
${\cal P}_{\gamma_E}$ are conjugate and hence the equatorial return times are
equal.  

Conversely, if the equatorial return times and angles are the same, then the
flows have the same frequency vectors (as functions of the global action angle
variables) and hence have the same global Birkhoff normal forms.  By Proposition
(1.5.2) the flows are symplectically equivalent.\qed

\section{Quantum dynamics and normal form}

It is owing to the following notion that simple surfaces of revolution are so
manageable.
\bigskip

\noindent{\bf (2.1)  Definition}~~~~{\it The wave group $e^{it \sqrt{\Delta}}$
of a compact, Riemannian n-manifold $(M,g)$  is {\em
quantum torus integrable} if there exists a unitary Fourier-Integral representation
$$\hat{\tau} : T^n \rightarrow U(L^2(M)),\;\;\;\;\;\;\hat{\tau}_{(t_1,\dots,t_n)} =
e^{i (t_1 \hat{I_1} + \dots t_n\hat{I_n})} $$
of the n-torus and a symbol $\hat{H} \in S^1(\Rr^n - 0)$ such that $\sqrt{\Delta} =
\hat{H}(I_1,\dots,I_n).$}
\bigskip

In the formula above,  we follow physics notation in indicating operators (as opposed
to their symbols) with a `hat'.  Thus, the generators $\hat{I}_j$ are first order
pseudodifferential operators with the property that $e^{2\pi i \hat{I}_j} = C_j Id$
for some constant $C_j$ of modulus one. 

Since $\hat{H}$ is a first order elliptic symbol on $\Rr^n - 0$ it 
 has an asymptotic expansion in homogeneous functions of the form:
$$\hat{H} \sim H_1 + H_o + H_{-1} + \dots, \;\;\;\;\;H_j( r I) = r^j
H_j(I).\leqno(2.2)$$
 The quantum action operators are  uniquely defined up to a choice of 
basis of $H^1(\Rr^n/\Z^n, \Z)$, 
 the terms $H_j$ are uniquely determined up to the same ambiguity.

The principal symbols $I_j$ of the $\hat{I}_j$'s generate a classical Hamiltonian
torus action, so any quantum torus action automatically induces a classical one. 
Conversely, it is a theorem of Boutet de Monvel-Guillemin and Weinstein that any
classical Hamiltonian torus action can be quantized [BM.G, Appendix, Proposition
6.6]. Since metrics in
${\cal R}$ give rise to quantum torus actions, we have, in particular:
\bigskip

\noindent{\bf (2.3) Proposition (cf. [CV.1])}~~~{\it Suppose $g \in {\cal R}$.  Then
$\sqrt{\Delta_g}$ is quantum torus integrable.}
\bigskip
 
We also observe that the following holds for any Laplacian commuting with 
a quantum torus action:
\bigskip

\noindent{\bf (2.4)  Proposition}~~~{\it For any $\Delta$,  $H_o = 0.$}
\bigskip

\noindent{\bf Proof:} Since the subprincipal symbol $\sigma_{sub}(\sqrt{\Delta})$
equals zero, the same is true $\hat{H}:= \hat{H}(\hat{I}_1, \hat{I}_1).$   Now
$\sigma_{sub}(\hat{H})$ is invariantly defined (independent of a choice of
symplectic coordinates); hence it may be expressed in action-angle coordinates in the
form 
$$0 = \sigma_{sub}(\hat{H}) = H_o - \frac{1}{i} \sum_j\frac{\partial^2
H_1}{\partial I_j
\partial \phi_j}. $$
But the mixed derivative term automatically vanishes since $H_1$ is a function only of
the classical action variables. \qed 

Let us specialize to the case of $\sqrt{\Delta_g}$ with $g \in {\cal R}.$  
 From the fact that $e^{2 \pi i \hat{I}_j} =
C_j Id$ for a quantum torus action, it follows that the joint spectrum of the
quantum moment map
$$Sp({\cal I}) \subset \Z^2 \cap \Gamma + \{\mu\}$$
is the set of integral   lattice points, translated by $\mu$, in the closed convex
conic subset $\Gamma \subset \Rr^2$. The vector $\mu = (\mu_1, \mu_2)$ can be
identified with the Maslov indices of the homology basis of the invariant torii.  In
the case of $\sqrt{\Delta_g}$, $\mu = (0, 1/2)$ [CV.1].
\medskip

\begin{figure}[thb]\label{figure5}
\centering{\epsfig{file= 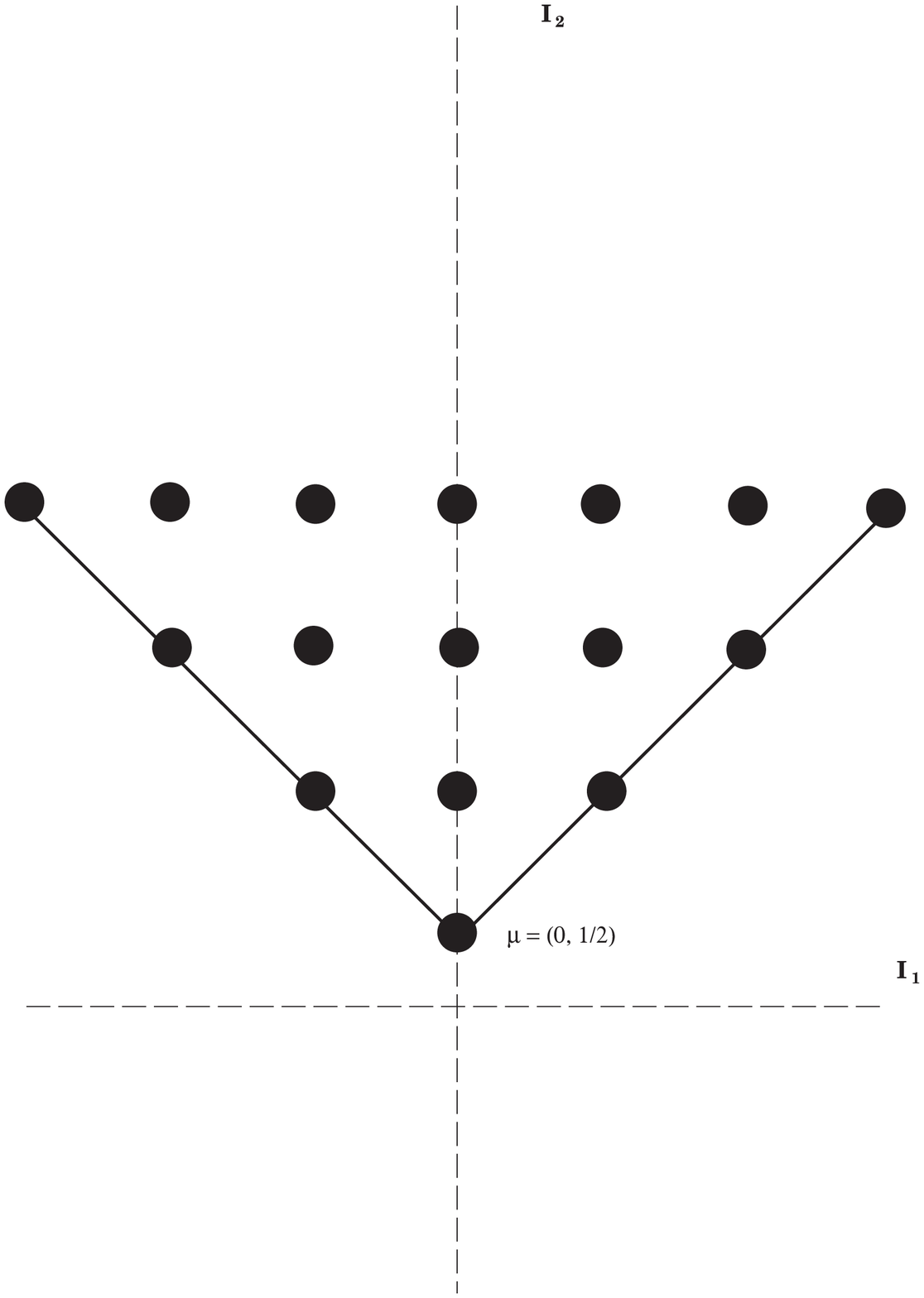, height= 3in}}
\caption{}
\end{figure}

\medskip

 Expressing
$\sqrt{\Delta_g}$ in the form $\hat{H}(\hat{I}_1,\hat{I}_2)$ we have that
$$Sp(\sqrt{\Delta_g}) = \{ \hat{H} ( N + \mu): N \in \Z^2 \cap \Gamma_o\}.$$
Thus, the  eigenvalues of $\sqrt{\Delta_g}$ have the form:
$$\lambda_N \sim H_1(N + \mu) + H_{-1}(N + \mu) + \dots$$
and the wave trace takes the form
$$Tr e^{it \sqrt{\Delta_g}} = \sum_{N \in \Z^2} e^{i t \hat{H}(N + \mu)}.
\leqno(2.5)$$

The symbol $\hat{H}$ may be regarded as a global
 quantum Birkhoff normal form. As in the classical case, it has a germ at any
periodic torus, so these may be regarded as the microlocal Birkhoff canonical
forms.  To be more precise, we imitate the definition of the classical tangential
Birhoff normal forms and write 
$$H_j(I_1, I_2) = (\omega^o \cdot I)^{j}\;\; H_j(\frac{I_1}{\omega^o \cdot I},
 \frac{I_2}{\omega^o\cdot I}) := (\omega^o\cdot I)^{j}\;\; h_j (\xi) \leqno(2.6)$$
where $\xi$ is a linear coordinate relative to a vector
$v$ generating $\omega^o \cdot I = 1.$
We then Taylor expand $h_j (\xi)$ around $\xi = 0$:
$$h_j(\xi) = \sum_{\alpha \geq 0} h_j^{\alpha}(0) \xi^{\alpha}.$$
\bigskip

\noindent{\bf (2.7) Definition}~~~{\it The homogeneous quantum Birkhoff normal form
coefficients of $\hat{H}$ at the periodic torus $T_{I^o}$ are the Taylor coefficients
$h_j^{\alpha}$ for $j = 1, 0, -1 ,\dots.$}
\bigskip

\noindent{\bf (2.8) Proposition}~~~{\it Assume that $g \in {\cal R}^*$.
 Then all of the functions $H_j$ are real analytic in the interior of
the cone $\Gamma_o$ and all of the functions $h_j$ are real analytic near $\xi = 0$.}
\bigskip

\noindent{\bf Proof}: First consider $|\xi|_g = H_1(I_1,I_2) = H(I_1, I_2).$ We
know that $H$ is a $C^{\infty}$ homogeneous function on $\Gamma_o$.  On the other
hand, from the formula $I_2 = G(|\xi|_g, I_1)$ we see that $I_2$ is a
real analytic function of $|\xi_g|, I_1.$  Since $G$ is the inverse function to
$H$ with respect to the first variable, $H$ must also be a real analytic function
of $I_1, I_2.$

Next recall that $\hat{I}_2$ is defined (up to a smoothing term) as the function 
$\hat{G}(\sqrt{\Delta}, \hat{I}_1)$ which has principal symbol $G(|\xi|_g, I_1)$ and
which has integral spectrum.  More precisely, one begins with $G(\sqrt{\Delta},
\hat{I}_1)$, which has the property that $exp( 2\pi i G(\sqrt{\Delta},
\hat{I}_1) + \mu_j) =  I + K$ with $K$ of order -1.  One defines $R =
\frac{-1}{2 \pi i} Log (I + K)$ and puts $\hat{G}(\sqrt{\Delta}, \hat{I}_1) = 
G(\sqrt{\Delta},\hat{I}_1) + \mu_j + R_j.$

Since $G$ is a real analytic function and we only apply the holomorphic functional
calculus in the steps of the construction, it follows that $\hat{G}$ is an analytic
function of $(\sqrt{\Delta},\hat{I}_1)$.

Since $\hat{I}_2 = \hat{G}(\sqrt{\Delta},\hat{I}_1)$ has the inverse function
$\sqrt{\Delta} = \hat{H}(\hat{I}_1, \hat{I}_2)$, it follows again by the inverse
function theorem for analytic functions that $\hat{H}$ is a real analytic function.

The real analyticity of the $h_j$'s follows from that of the $H_j$'s.
\qed
\bigskip

\section{Wave invariants as non-commutative residues}

To relate the wave invariants to the coefficient of the normal form,
 it will be  helpful (as in [G.1][Z.1,2]) to express the wave invariants as
non-commuative residues of the wave operator and its time-derivatives. Let
us recollect how this goes.

The non-commutative
residue of a  Fourier Integral operator is an extension of the well-known 
 non-commutative residue of a pseudodifferential operator 
$A$ on a compact manifold
$M$ , defined by 
$$\mbox{res}(A)=2\:\mbox{Res}_{s=0}\;\zeta (s,A)$$
where 
$$\zeta (s,A) = \:\mbox{Tr}\:A\Delta^{-s/2}\;\;(\mbox{Re }s>>0)\;.$$
Here, $\Delta$ is a Laplacian (or any positive elliptic operator) on $M$.
  From work of
Seeley, Wodzicki and Guillemin, one knows that  $\zeta(s,A)$ is
 holomorphic in Re $s>\half
\dim M +$ ord$(A)$ and admits a meromorphic continuation to $\C$,
 with simple poles at $s =
\dim M + $ord $(A) - k$ ($k=0,1,2,\ldots$).  The residue at $s=0$ has a number of
remarkable properties (not shared by the residues of the other poles):

\begin{tabbing}\noindent ~~~~~~\=--~~\=res$(AB) =$ res$(BA)$, i.e.\ res is a
trace on the algebra $\Psi^*(M)$ of pseudodifferential operators over $M$;\\

\>--\>res$(A)$ is independent of the choice of $\sqrt \Delta$;\\

\>--\>there is a local formula for the residue,
res$(A)=(2\pi)^{-n}\int_{S^{*}M}a_{-n}(x,\xi) i_{{\cal R}} dx \wedge d\xi $,
\end{tabbing}
where $a_{-n}$ is the term of degree $(-n)$ in the complete symbol expansion
$a\sim \displaystyle{\sum^{-\infty}_m a_j}$ for $A$; $dx \wedge d\xi $ is the
canonical symplectic volume measure on $T^*M$.

These results  may be extended to Fourier integral
operators as follows: Let
 $A$ be a Fourier Integral operator
 in $I^m(M\times M,\Lambda)$ for
some homogeneous canonical relation $\Lambda \subset T^*(M\times M)\backslash 0$ and
$m\in \Z$.   Below, diag$(X\times X)$ denotes the diagonal in $X\times X$.  
Below we will sketch a proof of the following:\bigskip 

\noindent {\bf (3.1) Theorem }~~{\it Suppose $\Lambda$ and diag$(T^* M\times T^*M)$
intersect cleanly.  Then $\zeta (s,A) =$ Tr $A\Delta^{-s/2}\;\;(\mbox{Res }>>0)$ has a
meromorphic continuation to $\C$, with simple poles at $s =
m+1+\frac{e_0-1}{2}-j$, where $e_0 = \dim \Lambda \cap$ diag$(S^*M\times
S^*M)$, and $j=0,1,2,\ldots$ .}\bigskip 

The clean intersection hypothesis above is that $\Lambda\cap$ diag$(T^*M\times
T^*M)\backslash 0$ is a clean intersection.  It is satisfied in
the case where $\Lambda = C_t$ if and only if the fixed point set of $G^t$ is clean. 
Hence, Theorem (3.1)  implies:\bigskip 

\noindent {\bf (3.2) Corollary}~~  {\it Let $\zeta (s,t) =$ Tr $U(t)
\Delta ^{-s/2}$.  

If the fixed point set of $G^{t}$ is clean, then $\zeta(\cdot, t)$
has a meromorphic continuation to $\C$, with simple poles at $s = 1 + \frac{\dim
S\mbox{ Fix}(G^t)-1}{2} - j$~~($j = 0, 1,\ldots,2$).  }
\medskip

Here, $S$ Fix$(G^t)$ is the set of unit vectors in the fixed point set of $G^t$.
In the case of a completely integrable system,  $S$ Fix$(G^L)$ is the union of
the periodic torii with period $L$.  We assume here, and henceforth, that the
periodic torii are all clean fixed point sets for $G^t$ on $S^*M$.

The non-commutative residue of the Fourier integral operator $A$ is then defined
by:
\medskip

\noindent{\bf (3.3) Definition}{\it 
$$ \mbox{res}(A):= \mbox{Res}_{s=0}\;\zeta(s,A)\;$$}
\smallskip
  
\noindent just as in the  case of pseudodifferential operators.  And just as in that
case, res$(A)$ has some remarkable properties:

- it is independent of the choice of $\Delta$;

- if either $A$ or $B$ is associated to a local canonical graph, then res$(AB) =$
res$(BA)$

- there is a local formula for res$(A)$.

The basic properties of res$(A)$ may be deduced from a singularity analysis of
the closely related distribution trace $S(t,A) :=$ Tr$ AU(t)$ (cf.  [Z.4]).
 Under the cleanliness
assumption above,  $S(t,A)$ is a
Lagrangean distribution on
$\Rr$ with singularities at the set of `sojourn times',
$${\cal S}{\cal T}=\{T:\exists (x, \xi) \in S^*M:(x, \xi, G^T (x, \xi))\in
\Lambda\}.$$
For a given sojourn time $T$, the corresponding set $W_T = \{(x, \xi):(x, \xi,
G^t(x, \xi))\in \Lambda\}$ of sojourn rays fills out a submanifold of $S^*M$ of
some dimension $e_T$.  With $N(\lambda ,A) = \sum_{\sqrt{\lambda _j}\leq
\lambda } (A\varphi_j, \varphi_j)$, we have \bigskip 

\noindent {\bf (3.4) ~~Proposition [Z.4, Proposition 1.10].}~~{\it If $\rho\in
C^{\infty}(\Rr)$ with $\hat \rho \in C^\infty_0(\Rr)$, supp $(\hat\rho)\cap {\cal
S}{\cal T} = \{0\}$ and $\hat\rho \equiv 1$ near 0, then
$$\rho*dN(\lambda ,A)\sim C_n\lambda ^{m+\frac{e_0-1}{2}}\sum\alpha_j\lambda
^{-j}\;,$$
where $C_n$ is a universal constant.  The coefficients have the form,
$$\alpha_j = \int_{\Lambda_{\Delta}}\omega_j d\lambda _\Delta$$
where $\Lambda_\Delta=\Lambda \cap$ diag$(S^*M\times S^*M)$,
$e_0=\dim\Lambda_\Delta$, $d\lambda _\Delta$ is a canonical density on
$\Lambda_\Delta$ and the functions $\omega_j$ are determined by the $j$-jet of
the (local) complete symbols of $AU$ along $\Lambda_\Delta$.}\bigskip

\noindent {\it Proof of Theorem (3.1).}~~As in the pseudodifferential case [DG,
Proposition 2.1], we have
$$\begin{array}{lll}\mbox{Tr}\:A\Delta^{-s/2} &=&\langle \chi_s, dN(
^{\mbox{{\Large{\bf .}}}},
A)\rangle\\
&=&\langle \chi_s, \rho*dN(^{\mbox{{\Large{\bf .}}}}, A)
\rangle + \mbox{ entire}.\end{array}$$

Hence, 
$$\begin{array}{lll}\zeta (s,A)&=&C_n\displaystyle{\sum^\infty_{j=0} \alpha_j
\int^\infty_1\lambda ^{\frac{e_0-1}{2}+m-s-j} d\lambda +\mbox{ entire}}\\[14pt]

&=&C_n\displaystyle{\sum^\infty_{j=0} \frac{\alpha_j}{m+1+\frac{e_0-1}{2}-(s+j)}
+\mbox{ entire}}\;,\end{array}$$
completing the proof.~~\rule{2mm}{4mm}\bigskip 

In particular, if $A = U(L)$, then $\{0\}$ is a sojourn time if and only if $
L \in \mbox{ Lsp}(M, g)$.  If $L\not\in$ Lsp$(M,g)$, $\zeta (s, L)$ is regular
at 0.  

Now let us return to the case where  the geodesic flow is completely integrable. 
Then the dimension of each periodic torus ${\cal T}$ of period $L$ equals $e_o =
dim {\cal T} = n$.  Hence we have:
$$Tr U(t) = e_o(t) + \sum_{{\cal T}} e_{{\cal T}}(t) \leqno(3.5a)$$
where the sum runs over the periodic tori in $S^*M$ and where
$$e_{{\cal T}} (t) = a_{{\cal T}; - \frac{n+1}{2}} (t - L + i0)^{- \frac{n+1}{2}}
+ a_{{\cal T}; - \frac{n+1}{2} + 1} (t - L + i0)^{- \frac{n+1}{2} + 1 } +
\dots.\leqno(3.5b)$$ More precisely, it takes this form if $n$ is even; if $n$ is odd,
the positive  powers of $(t - L + i0)$ should be multiplied by $log (t - L + i0)$.
In the following Corollarly we use the notation $- {\cal T}$ for the time reverse
torus $\sigma {\cal T}.$
\bigskip

\noindent{\bf (3.6) Corollary}{\it 
$$\sum_{\pm} a_{\pm {\cal T}, -(\frac{n+1}{2}) + k} =
\:\mbox{res}(\sqrt{\Delta}^{-\frac{n+1}{2} + k} U(t)|_{t=L})$$} \bigskip 

\noindent {\it Proof.}~~In the notation of (possibly negative) fractional derivatives
in $t$, we have
 $\Delta^{\mu} U(L) = D^{\mu}_tU(t)|_{t=L}=$. The claim follows from the facts that
$$D^{-\frac{n+1}{2} + k}_t (t - L + i0)^{-\frac{n+1}{2} + k}= log (t - L + i0)$$
and that the non-commutative residue is the coefficient of $log (t - L + i0)$.\qed
\bigskip

\noindent{\bf (3.7) Examples}:
\bigskip

\noindent(a) If $dim M = 2$, then the wave trace expansion at a torus ${\cal T}$
has the form
$$e_{{\cal T}}(t) = a_{{\cal T},-\frac{3}{2}} (t - L + i0)^{- \frac{3}{2}} +
a_{{\cal T},-\half} (t - L + i0)^{- \frac{1}{2}} + \dots.$$
Hence
$$\sum_{\pm} a_{\pm {\cal T}, -\frac{3}{2}}= res \sqrt{\Delta}^{- \frac{3}{2}} e^{i L
\sqrt{\Delta}},
\;\;\;\;\;a_{{\cal T},-\frac{1}{2}} = res \sqrt{\Delta}^{ -\frac{1}{2}} e^{i L
\sqrt{\Delta}},
\dots$$

\noindent(b) If $dim M = 3$,  the wave trace expansion at  ${\cal T}$
has the form
$$e_{{\cal T}}(t) = a_{{\cal T},-2} (t - L + i0)^{- 2} + a_{{\cal T},-1} (t - L
+ i0)^{-1} + a_{{\cal T},0}log (t - L + i0) + \dots.$$
Hence
$$\sum_{\pm} a_{\pm {\cal T},-2}= res \Delta^{- 2} e^{i L
\sqrt{\Delta}},\;\;\;\;\;a_{{\cal T},-1}= res
\Delta^{- 1} e^{i L \sqrt{\Delta}},
\;\;\;\;\;\sum_{\pm} a_{\pm {\cal T},0} = res  e^{i L \sqrt{\Delta}},
\dots.$$ 

\noindent{\bf (3.8) Remark}: The wave invariants for a closed geodesic
$\gamma$ (or periodic torus $T_L$) are exactly the same as for its time
reversal, hence the same residue formulae also give the individual wave invariants.
For this reason it is often not necessary to resolve the ambiguity between 
a torus and its time reversal.

\section{Proof of the Main Lemma}

We now prove that the wave
trace invariants of a metric $g \in {\cal R}^*$ determine its quantum normal form
 $\hat{H}$. 

\subsection{ Wave invariants for simple surfaces of revolution}

We begin by specializing Corollary (3.6) to the case of a surface of revolution
$(S^2,g)$ in ${\cal R}^*.$   Since the length spectrum ${\cal L}$ is simple, there is
a unique periodic torus ${\cal T}_L \subset S^*_g S^2$ of each length $L \in {\cal
L}$  (up to time reversal).
 By the existence of global action-angle coordinates, it may be expressed
 in the form  ${\cal T}_L = T_{I_L}$ for a unique point $I_L = (I_{L1}, I_{L2}) \in
{\cal P}$ (up to reflection).  Let $\omega_L$ denote the frequency vector at $I_{L}$.
Then we have
$L \cdot \omega_L = M_L$ where $M_L$ is the vector of winding numbers of the periodic
orbits on ${\cal T}_L.$  

The other periodic torii of period $L$ lie on the rays $ \Rr^+ {\cal T}_L
\cup \Rr^+ \sigma({\cal T}_L)$.  Their
action coordinates lie on the rays $\Rr^+ I_L \cup \Rr^+ \sigma(I_L)$ and they have
the same frequency vector, $\omega_L$, as for $I_L$. Since the wave invariants at
${\cal T}_L$ depend only on the microlocalization of the wave group to a conic
neighborhood of
$\Rr^+{\cal T}_L$, we introduce a microlocal cut off operator $\hat{\psi}_L(\hat{I}_1,
\hat{I}_2)$
 with $\psi$ homogeneous of
degree 0, equal to 1 in a small conic  neighborhood of the ray $\Rr^+ I_L$ and zero
off of a slightly larger conic neighborhood. The singularity of $Tr
U(t)$ at
$t=L$ is then the same as the singularity of $Tr \hat{\psi}_L~~ U(t).$   

To calculate the wave trace as a residue, we also introduce the first order
pseudodifferential operator $\omega_L \cdot \hat{I} := \omega_{L1}\hat{I}_1
 + \omega_{L2} \hat{I}_2$. We emphasize that $\omega_L \cdot \hat{I}$ is a linear
combination with constant coefficients of the action operators.
   Since $\omega_I \cdot I = H$, the
principal symbol  $\omega_L \cdot I$ of $\omega_L \cdot \hat{I}$  takes the
value 1 at $I = I_L$ and therefore
$\omega_L \cdot
\hat{I}$ is elliptic in a conic neighborhood of ${\cal T}_L$.  We will use it
as the gauging elliptic operator in the residue formula for the wave invariants.
\bigskip

\noindent{\bf (4.1.1)~~ Proposition}~~~{\it Let $g \in {\cal R}^*$
and let  $L \in {\cal L}$.  Then we have:
$$\sum_{\pm} a_{\pm T_L, -\frac{1}{2} + k} =  Res_{s=0} \int_{\Gamma}
\psi_L (I +
\mu) e^{i 
\langle M_L, I \rangle} e^{-i L \hat{H}(I + \mu)}  (\hat{H}(I + \mu))^{- \frac{
1}{2} + k} (\omega_L \cdot (I + \mu))^{-s}dI$$
where as above $M_L$ is the vector of winding numbers of ${\cal T}_L$.}
\medskip

\noindent{\bf Proof}: Since $\sqrt{\Delta} = \hat{H}(\hat{I}_1, \hat{I}_2)$ and
since $\hat{\psi}_L$ is a function of the action operators, we have by Corollary (3.6)
that
$$ \sum_{\pm} a_{\pm T_L, -\frac{1}{2} + k} = Res_{s=0} \sum_{N \in \Z^2} \psi_L (N +
\mu) e^{i L
\hat{H}(N + \mu)} (\hat{H}(N + \mu))^{ - \frac{1}{2} + k} (\omega_L \cdot
(N + \mu))^{-s}.$$
We then apply the Poisson summation formula for Re $s >>0$ to replace the sum over $N
\in
\Z^2$ by 
$$Res_{s=0} \sum_{M \in \Z^2} J_{L, M, k} (s)$$
where
$$J_{L, M, k} (s):= \int_{\Gamma} \psi_L (I + \mu) e^{-i  \langle M,
I \rangle} e^{i L \hat{H}(I + \mu)} 
 (\hat{H}(I + \mu))^{ - \frac{1}{2} + k} (\omega_L \cdot (I + \mu))^{-s}dI.$$
It follows from Theorem 3.1 and by simplicity of the length spectrum that only
the term with $M = M_L$ has a pole at $s = 0.$  This can be seen more directly
from the fact that only in this term does the phase  
$- \langle M, I \rangle + L H(I)$  have a critical point, since
$M = L \nabla_I H (I_M)$ implies that the torus with actions $I_M$ is periodic
of period $L$ (cf. \S 1.2). 
\qed
\medskip

To calculate the residue of the  integral $J_{L k}(s):= J_{L, M_L, k} (s)$, we
rewrite the integrals in terms of the 
$(\rho, \xi)$ coordinates introduced in (1.4.3) 
in a conic neighborhood of the point $I_L \in \{H = 1\}$:
Recall that we parametrize the tangent line $T_{I_L} (\{H=1\})$ by
$$\xi \in \Rr \rightarrow I_L + \xi v$$
where $v$ is a non-zero vector along the line $M_L \cdot I = 0$ and
parametrize a conic neighborhood of $\Rr^+ I_L$ by
$$ (\rho, \xi) \in \Rr^+ \times \Rr \rightarrow \rho (I_L + \xi v).$$
Since $I_L$ is fixed, we abbreviate $h_{I_L}$ by $h$ in the next proposition.
We also denote the cutoff in these coordinates by $\psi_L(\xi).$
\bigskip

\noindent{\bf (4.1.2)~~ Proposition}~~~{\it With the above notation, 
$a_{T_L, -\frac{1}{2} + k}$ equals the term of order $\rho^{-1 - k}$
 in the asymptotic
expansion of the integral
$$J_{L k}(\rho):= L^{-2 + s - k + \half} e^{i \langle M, \mu \rangle} \rho^{\half}
\int_{\Rr} \psi_L(\xi) e^{i  \rho \alpha
\frac{\xi^2}{2}} e^{i  \rho g_3} e^{i \sum_{k=1}^{\infty} L^{k + 1} \rho^{-k}
h_{-k}(\xi)} [ h(\xi) +
\sum_{\ell = 1} ^{\infty} L^{\ell + 1} \rho^{-\ell - 1} h_{\ell} (\xi)]^{ -
\frac{1}{2} + k} d\xi.$$
\noindent Here, $\alpha =  h''(0)$, and $g_3$ is the third order remainder in the
Taylor expansion of $h(\xi)$ at $\xi = 0.$}
\bigskip

\noindent{\bf Proof} We first change variables 
 $I + \mu \rightarrow I$ in the expression for $J_{ L, M_L, k}(s)$ in the preceding
Lemma and then further change variables to   $(\rho := \omega_M \cdot I,
\xi)$.  Thus we get
$$J_{L k}(s) = e^{i \langle M, \mu \rangle} \int_{o}^{\infty} \int_{\Rr} \psi_L(\xi)
e^{-i \rho \langle M, (I_L +  \xi v \rangle} e^{i L [ \rho h(\xi) + \rho^{-1}
h_{-1}(\xi) + \dots]} [ \rho h(\xi) + \rho^{-1} h_{-1}(\xi) + \dots]^{-\half + k}
\rho^{-s+1} d\rho d\xi.$$ Taylor expanding $h(\xi) = h(0) + h'(0) \xi + \half h''(0)
\xi^2 + g_3 (\xi)$ and using that $h(0) = H(I_L) = 1$, that $h'(0)= \omega_{I_L}
\cdot v = 0$ and that $\langle M, I_L + \xi v \rangle = \langle M, I_L \rangle = L$
by (1.2.1a-b), we get  
$$J_{L k}(s) = e^{i \langle M, \mu \rangle} \int_{o}^{\infty} \int_{\Rr} \psi_L(\xi) 
e^{i L  \rho h''(0) \half \xi^2} e^{i L \rho g_3(\xi)} e^{iL [
\rho^{-1} h_{-1}(\xi) + \dots]} [ \rho h(\xi) + \rho^{-1} h_{-1}(\xi) +
\dots]^{-\half + k} \rho^{-s+1} d\rho d\xi.$$

Now change variables again, $\rho \rightarrow L^{-1} \rho$ and pull the factor
$\rho L^{-1}$ in front of $h(\xi)$ in the bracketed expression outside the $d\xi$
-integral.  We get:
$$J_{L k}(s) = e^{i \langle M, \mu \rangle} L^{-2 + s - k + \half}
 \int_{o}^{\infty}\{ \int_{\Rr}
\psi_L(\xi)  e^{i   \rho h''(0) \half \xi^2} e^{i  \rho g_3(\xi)} e^{iL [
L \rho^{-1} h_{-1}(\xi) + \dots]}$$
$$ [  h(\xi) + \rho^{-2}L^{2} h_{-1}(\xi) +
\dots]^{-\half + k} d\xi\} \rho^{-s+1 + k - \half} d\rho .$$
The pole at $s=0$ is produced by the terms of order  $\rho^{-1}$ in the
$d \xi - d \rho$-integrals, hence by the terms of order
 $-k-1 - \half$ in the 
asymptotic expansion of the $d \xi$-integral. \qed
\bigskip

To determine the terms of order $\rho^{-k-1}$ in $J_{L k}(\rho)$ we apply the
method of stationary phase. Cancelling factors of $\rho^{\half}$ we get:
\bigskip

\noindent{\bf (4.1.3)~~ Corollary}~~~{\it In the above notation, 
$a_{T_L, -\frac{1}{2} + k}$ equals the term of order $\rho^{-k - 1}$ in the asymptotic
expansion of
$$\frac{1}{\sqrt{2 \pi i \alpha}}L^{-2   - k + \half} e^{i \langle M, \mu
\rangle}\sum_{ m = 0}^{\infty} (2 i\rho)^{-m}
 \alpha^{-m}
\partial_{\xi}^{2 m} \{e^{i 
\rho g_3(\xi)} e^{i \sum_{j=1}^{\infty} L^{j + 1}
\rho^{-j} h_{-j}(\xi)} [h(\xi) +
\sum_{\ell = 1} ^{\infty} L^{\ell + 1} \rho^{-\ell - 1} h_{- \ell} (\xi)]^{ -
\frac{1}{2} +  k}\}|_{\xi = 0} $$}
\bigskip

\subsection{The meridian torus}

We now want to analyse the terms of the  expansion coming from the ray of  
meridian torii with $I_1 = 0.$  The special property of the meridian torii
is that they are invariant under  the canonical  involution 
$\sigma(x,\xi) = (x, -\xi)$ of $T^*S^2-0$. As the following proposition shows,
this leads to a useful symmetry property of the quantum normal form. 
  As above, the notation $h(\xi), h_{-1}(\xi)$
etc. refer to the functions $h_{I_L}$ (etc.) where   $I_L$ is the point on $\{H=1\}$
corresponding to the meridian torus in $S^*_g S^2$.
\bigskip

\noindent{\bf (4.2.1)~~ Proposition}~~~{\it The complete symbol of $\hat{H}$ is
invariant under $\sigma$. Hence $h_j$ is even for all j.}
\bigskip

\noindent{\bf Proof}~~ Let $C$ denote the operator of complex conjugation:
$C \psi = \bar \psi.$  Since $\sqrt{\Delta}$ commutes with $C$, so does
$\hat{H}(\hat{I}).$  Since $C$ is a conjugate-linear involution, this implies that
$$\hat{H}(\hat{I}) = C^{-1}  \hat{H}(\hat{I}) C = \bar{\hat{H}}(C^{-1}\hat{I} C)$$
where the bar denotes complex conjugation.

We claim that $C^{-1} \hat{I}_1 C = - \hat{I}_1$ and that $C^{-1} \hat{I}_2 C = 
\hat{I}_1$. Moreover, that $I_1 \circ \sigma = -I_1, I_2 \circ \sigma = I_2.$
The  statements regarding $I_1, \hat{I}_1$ are obvious since
$\hat{I}_1 = \frac{\partial}{i \partial \theta}$ changes sign under complex
conjugation and since $\sigma_{C^{-1} \hat{I}_1 C} = I_1 \cdot \sigma.$  
  This latter
also follows from the fact $I_1(x,\xi) = x_2 \xi_1 - x_1 \xi_2$.

Regarding $I_2$ we note that its $\sigma$-invariance follows immediately 
 from the explicit formula (1.3.1).
The non-obvious claim is that $\hat{I}_2$ is invariant under $C$. But from the
invariance of the principal symbol we have
$$C^{-1} \hat{I}_2 C = \hat{I}_2 + K_1$$
where $K_1$ is of order 0. Since $\hat{I}_2$ is a function of $(\hat{I}_1,
\sqrt{\Delta})$, it is clear that $C$ must take joint $(\hat{I}_1, \hat{I}_2)$
eigenfunctions into joint eigenfunctions.  
Hence $C$ determines an involution on the joint spectrum $\{(n, k+ \half): |n| \leq
2k+1, k\geq 0\}$ which, we recall, is simple. Consequently, the
involution (still denoted $C$) must take the form
$$C : (n, k+ \half) \rightarrow (-n, k + \half + f(n,k))$$
where $f$ is a bounded  function. Moreover, since $f(n,k) = \langle \hat{I}_2
\overline{\phi_{n,k}}, \overline{\phi_{n,k}}\rangle$ and since the $\phi_{n,k}$
are quasi-modes associated to Bohr-Sommerfeld-Maslov torii [CV.3], it is clear
that $f(n, k + \half)$ must be a polyhomogeneous function of order 0 on $\Gamma_o$.
Since it is also integral-valued on the semi-lattice of joint spectral points, it must
be constant. Additionally it must satisfy the involution condition, and one sees
that the constant must be 0. 

Returning to $\hat{H}(\hat{I}_1, \hat{I}_2)$ we have $ \hat{H}(\hat{I}_1,
\hat{I}_2)  = C^{-1} \hat{H}(\hat{I}_1, \hat{I}_2) C  =
\bar{\hat{H}}(C^{-1} \hat{I}_1C,C^{-1}\hat{I}_2C) =
\bar{\hat{H}}(-\hat{I}_1,\bar{I}_2)$ so that
$\hat{H}(a,b) =
\bar{\hat {H}}(-a,b).$  We then observe that $\hat{H}$ is a real function (at least
modulo terms of order $- \infty.$) To see this, we recall that the eigenvalues
$\hat{H}(n, k+\half)$ are real and argue by induction on the symbol expansion using
that 
$$\hat{H}(n, k+ \half) = H(n,  k + \half) + H_{-1}(n, k + \half) + \dots \in \Rr.$$ 
First, the principal symbol $H$ is real so we may drop it from the expansion
without affecting reality.  Assuming that $H = H_1, H_{-1} ,\dots, H_{-k}$ are
real, we may drop them all and get that the tail sum is real.  Since it is
dominated by $H_{-k-1}$, this function must be real for all points
$(n, k+ \half)$ sufficiently far from $0$.  But by homogeneity, $H_{-k-1}$ is then
real on the projection of these lattice points to $H = 1.$  The projected points
form a dense set by Proposition 1.2.4  and hence $H_{-k-1}$ is everywhere real. We
conclude that
$\hat{H}$ is $\sigma$-invariant.

Now consider the ray of meridian torii, or more precisely the ray
$\Rr ^+ (0,1)$ in the action cone $\Gamma.$   We note that this ray is invariant
under the involution $\sigma_{\Gamma}(a,b):= (-a,b)$ of $\Gamma$, and
moreover the meridian torus is invariant under $\sigma$ (it `rotates' the torus by
angle $\pi.$) From the above, the level set $\{H=1\}$ is invariant under  
$\sigma_{\Gamma}$ and hence the tangent line at $(0,1)$ is invariant.  Evidently
it is horizontal in the $(I_1, I_2)$-plane and $\sigma_{\Gamma}$ restricts to it
to the map $\xi \rightarrow - \xi.$  Since the complete symbol of $\hat{H}$ is
$\sigma_{\Gamma}$-invariant, the $h_j$'s must be even. \qed
\bigskip

We now go back to the wave invariants associated to the meridian torus and
its iterates.  Let us write the amplitude for the kth wave invariant of the
 iterate of length $L$, namely 
$$A_{Lk}(\xi,\rho):= e^{i \sum_{j=1}^{\infty} L^{j + 1}
\rho^{-j} h_{-j}(\xi)} [h(\xi) +
\sum_{\ell = 1} ^{\infty} L^{\ell + 1} \rho^{-\ell - 1} h_{- \ell} (\xi)]^{ -
\frac{1}{2} +  k},\leqno(4.2.2)$$
in the form
$$A_{Lk}(\xi,\rho): = A_{Lko}(\xi) + \rho^{-1} A_{L k 1}(\xi) + \dots.$$
Thus, $a_{T_L, -\half + k}$ is the term of order $\rho^{-k-1}$ in
$$\frac{L^{-k-1}}{\sqrt{2 \pi i L \alpha}} \sum_{m=0}^{\infty}\sum_{\ell =0}^{\infty}
\alpha^{-m} (2 i\rho)^{-m - \ell}
\partial_{\xi}^{2m} [e^{i \rho g_3} A_{L k \ell}].\leqno(4.2.3)$$

Expanding the derivatives and using that $g_3 (\xi)$ is even and of order 
$0(\xi^4)$, we may rewrite (4.2.3) in the form
$$\frac{L^{-k-1}}{\sqrt{2 \pi i L \alpha}} \sum_{m=0}^{\infty}\sum_{\ell =0}^{\infty}
\sum_{j=0}^{m} \sum_{q \leq \half m} \sum_{(i_1, \dots, i_q): |i| = j} \{
C_{mli} \alpha^{-m}
\rho^{-m -\ell + q} (\partial_{\xi}^{2 i_1}g_3 \dots
\partial_{\xi}^{2 i_q}g_3) \partial_{\xi}^{2(m-j)} A_{L k \ell}\}.\leqno(4.2.4)$$
Here, the multi-index $i$-sum runs over $q$-tuplets with $i_n \geq 2, |i| = \sum
i_n = j \leq m$ with $q \leq \half m, 2q \leq |i| = j \leq m  .$ 
In order that $-m - \ell + q = -k - 1$ it is necessary that
 $$m \leq 2(k + 1),\;\;\;\;\;\;\;\;\;\;\ell \leq k + 1.$$ 
 These restrictions follow from the fact that it takes at least 4
derivatives of $g_3$ to make a non-zero contribution.

\subsection{A collection of formulae}

For future reference
we assemble some notation and identities  regarding the coefficients $A_{Lk \ell}$ and their relations
to the wave invariants.

\bigskip

\noindent{\bf (4.3.1) Notation}
\bigskip

\begin{tabular}{l}
(1) Define: $F_n(h_{-1}, \dots, h_{-n})$ by $[h(\xi) + \sum_{\ell = 1} ^{\infty}
L^{\ell + 1} \rho^{-\ell - 1} h_{- \ell} (\xi)]^{ - \frac{1}{2}} =$\\
$h(\xi)^{ -
\frac{1}{2}}[ 1 + \sum_{n=2}^{\infty} L^n \rho^{-n}F_n(h^{-1}, h_{-1},
\dots, h_{-n})]  $\\
 \\ \hline
(2) Define: $G_n(L,h_{-1}, \dots, h_{-n})$ by $e^{i \sum_{j=1}^{\infty}
 L^{j+1} \rho^{-j} h_{-j}} = 
 1 + \sum_{n=1}^{\infty} L^{n} \rho^{-n} G_n(L, h_{-1}, \dots, h_{-n}).$\\ 
\\ \hline
(3) Define $F_{nmj}$ by $F_n(h^{-1}, h_{-1}, \dots, h_{-n}) = \sum_{m = 1}^n h^{-m}
[\sum_{j =(j_1,
\dots, j_m) : |j| = n} F_{nmj} h_{-j_1}
\dots h_{-j_m}] $ \\ 
\\ \hline
(4) Define $G_{nmj}$ by $G_n(L,h_{-1}, \dots, h_{-n}) = 
\sum_{m=1}^n L^m [\sum_{j= (j_1, \dots, j_m): |j| = n} G_{nmj}   h_{-j_1} \dots
h_{-j_m}] $ \\ 
 \\ \hline

\end{tabular}
\bigskip

\noindent{\bf (4.3.2) Identities}
\bigskip

\begin{tabular}{l}

(1)  $A_{L0n}$ =~~~  $ L^n h^{-\half} [F_n + G_n + \sum_{i + j = n} F_i G_j]$ \\ 
 \\  \hline
(2)  $A_{Lk}$ =~~~  $ [h(\xi) +
\sum_{\ell = 1} ^{\infty} L^{\ell + 1}
 \rho^{-\ell - 1} h_{- \ell} (\xi)]^{  k} A_{L0}.$\\
\\ \hline
(3) $A_{L0n} = L^{n}h^{-\half} [\sum_{m=1}^n \sum_{j=(j_1,\dots, j_m): |j| = n}
\{F_{nmj}h^{-m} + G_{nmj}L^m\} h_{-j_1}\dots h_{-j_m}$\\
\\ \vspace{.3in} $+ \sum_{a + b = n} \sum_{m_1 = 1}^a \sum_{m_2 = 1}^b \sum_{i=(i_1,
\dots, i_{m_1}): |i|=a}
\sum_{j=(j_1, \dots, j_{m_2}): |j|=b}F_{am_1i}G_{bm_2j} h_{-i_1}\dots h_{-i_{m_1}}
 h_{-j_1} \dots h_{-j_{m_2}}].$\\  \hline
 (4)  $A_{Lk\ell} = h A_{L(k-1)\ell} + \sum_{ij: i + j + 1 = \ell, j\geq 1} L^{j+1}
h_{-j} A_{L k i}.$\\
\hline
\end{tabular}
\bigskip

The coefficients $F_{nmj}, G_{nmj}$ are universal and hence up to the prefactor
of $h^{-\half}$, 
$A_{Lkj}$ is a (non-homogeneous) polynomial of
degree $j$ in $L$,  in  $ h^{-1},$ and in the $h_{-j}$'s with universal coefficients. 
 In particular, the first few $A_{Lkj}$'s are given by:
$$A_{Lko} = h(\xi)^{ -\frac{1}{2} +  k},\;\;\;\;\; A_{L k 1} =
h(\xi)^{ -\frac{1}{2} +  k} L^2 h_{-1},\;\;\;\;\;\; A_{Lk2} = 
h(\xi)^{ -\frac{1}{2} +  k}[
L^3 h_{-2} +  L^{4} h_{-1}^2 + (k - \half)  L^2 \frac{h_{-1}}{h}].$$

It follows that the principal wave invariant is given by 
$$a_{T_L, -\frac{3}{2}}:= c_L= \frac{1}{\sqrt{2 \pi i \alpha L}}
e^{i \langle M_L, \mu \rangle}$$
and that the higher
wave invariants $a_{T_L, -\half + k}$ are given by $c_L$ times 
polynomials
in
$L$ and in the derivatives of  $h, h^{-1}, h_{-1}, \dots$ at $\xi = 0$.  For
instance, the subprincipal wave invariant in dimension 2 
 is given given in terms of universal coefficients $C'_{ijk}$ by:
$$a_{T_L, -\half} = c_L [C'_{004}\partial_{\xi}^4 g_3 A_{L00}  + C'_{010} A_{L01} +
C'_{002} \partial_{\xi}^2 A_{L00}]$$
which is easily seen to equal
 $$ c_L [ C_{004} \partial_{\xi}^4
h(\xi)|_{\xi = 0} + C_{010} L^2 h_{-1}(0) +C_{002} \partial_{\xi}^2
h(\xi)^{-\half}|_{\xi = 0}].  $$

\bigskip

Above the indices in the coefficients $C_{kmj_1j_2\dots}$ have the following meaning:
$k$ corresponds of
course to the $k$ index; $m$ is the power of $L$ and $j_n$
 are the jet orders of $h_{-n}$, with
with exception of $j_o$ which is the jet order of $h$.  
  \bigskip

\noindent{\bf (4.3.3)~~ Proposition}~~~{\it We have

 $$a_{L, -\half + k} = c_L
P_k(L, h^{(2)}(0), \dots, h^{(2k + 4)}(0), h_{-1}(0), \dots,
 h_{-1}^{(2k)}(0),
\dots, h_{-k}(0), h_{-k}^2(0), h_{-k-1}(0))$$
where $P_k$ is a polynomial with the following properties:

\noindent (i) It involves only the first
$2k+4$ Taylor coefficients of $h$ at $0$, the first $2k$ of $h_{-1}$,..., the 
first $2k + 2 - 2n$ of $h_{-n}$ ...,
the first $2$  of $h_{-k}$ and the 0th of $h_{-k-1}.$

\noindent(ii) It is of degree 1 in the variables $h_{-k-1}(0)$,
 $h_{-k}^{(2)}(0), \dots, h^{(2k + 4)}(0)$, and
each occurs in precisely one term.

\noindent(iii) The $L$-order of the monomials containing these 
terms is respectively $L^{k+2}, L^{k+1},\dots,
L^0.$ }
\bigskip

\noindent{\bf Proof}:  To prove these claims we combine the formula for
$a_{T_L, -k + \half}$ in (4.2.4) with the formulae for the $A_{Lk\ell}$ given
in (4.3.2).   Since we may replace the factors of $g_3$ by $h$ in (4.2.4) and
since (4.3.2) expresses $A_{Lk \ell}$ as $h^{-\half}$ times a polynomial in $L$
and in the $h_{-j}$'s, it is clear that $a_{T_L, -k + \half}$ is given by
 a polynomial in the data stated above. It remains to prove that the polynomial
has the properties claimed in (i)--(iii).

 We prove these claims by proving the stronger statement 
that  $$a_{L, -\half + k}
 = c_L Q_k(L, h(\xi)^{-\half}, h(\xi), h^{(2)}(\xi)), \dots, h^{(2k +
4)}(\xi)), h_{-1}(\xi)), \dots, h_{-1}^{(2k)}(\xi)), \dots,
 h_{-k}(\xi)), h_{-k}^2(\xi)), h_{-k-1}(\xi)))|_{\xi
= 0}$$ 
where $Q_k$ has the properties (i)-(iii) for variable $\xi.$

The proof is by induction on $k$.  The properties are 
 visibly true for the principal and subprincipal 
wave invariants. Assume then that they are correct for $k \leq N$ and consider how
things change as $N \rightarrow N + 1$.   First, 
the amplitude $A_{L (N+1)}$ is given by 
 $[h + L^2\rho^{-2} h_{-1} + \dots]A_{L N}$. Second,   we are looking at the
term of order $\rho^{-N-2}$ rather than $\rho^{-N-1}$ in
 the asymptotic series (4.2.4).

In going one further order into the asymptotic series, only two new things
happen:

\noindent $\bullet$ The term $h_{-N-2}$ appears for
the first time, arising from the linear term in the Taylor expansion of
the exponential in (4.2.2).  The linear term in the binomial expansion of
the power in (4.2.2) does not contribute at this stage because its $\rho^{-1}$
factor has one higher power.

\noindent $\bullet$   
 Two additional derivatives are allowed to fall on  the previous $h_j$'s.  A priori,
there could be between two and six additional derivatives in a method of stationary
phase expansion.  However, the cases of three to six derivatives do not contribute
any new data.  Indeed, the cases of three-four derivatives occur when $m$ goes up
by one.  One has to remove the extra factor of $\rho^{-1}$ by applying at least
one derivative to the $e^{i \rho g_3}$ factor.  But in fact all four have to be
applied to $g_3$ to make a non-zero contribution, and hence no derivatives are left
to apply to the $h_{-\ell}$'s.  In the case of five or six derivatives, where $m$
goes up to two, one
needs to remove two extra factors of $\rho^{-1}$ by applying derivatives to the $e^{i
\rho g_3}$ factor, and there is no non-zero way to do this.

 Claim (i)
follows immediately from these observations. 

To prove (ii) we note that by the induction hypothesis, $h_{-N-1}(\xi), \dots, 
h^{(2N + 4)}(\xi)$ occur linearly
in $P_{N}(\xi).$  Since it requires both of 
the two new derivatives to fall on these factors to produce
$h_{-N -1}^{(2)}$ etc. in $P_{N+1}(\xi)$, these factors 
will also occur to order 1.  As for $h_{-N-2}$, we observed above that it comes only
from the linear term in the exponential in (4.2.2) and hence it too appears to order
1.

The proof of (iii) follows the proof of (ii).  Since the terms in the exponent
of the exponential factor in (4.2.2) have the form $L^{j+1} \rho^{-j}$,
 the new $h_{-N-2}(\xi)$ term has the coefficient $L^{N+3}$. 
Similarly, the other terms under discussion, e.g. $h_{-N - 2 + r}^{(2r)}(\xi)$, 
originated as $h_{-N - 2 + r}$ at the $N + 1 -r$th 
stage with the factor of $L^{N +3 -r}$. We would like to show that they remain with
just this factor of $L$ as we move inductively up to the $N+1$st stage.
To this end, we note that in order to produce the term $h_{-N-2-r}^{(2r)}$ 
  at the (N+1)st stage, $2r$  derivatives must fall
on the original $h_{-N - 2 + r}$. But there are exactly $r$ stages intermediate
between the $N+1 -r$th stage and the $N+1$st stage and at each stage at most
two new derivatives can fall on a factor.  Hence, each of the two new derivatives at
 each stage must fall on the factor of concern. 

Let us also consider what can be multiplied against the original $h_{-N - 2 + r}$
in the course of producing $h_{-N - 2 + r}^{(2r)}(\xi)$.  We observe that
 each  new pair of
derivatives is accompanied by a factor of $\rho^{-1}.$  
 Since  $h_{-N-2 + r}(\xi)$,  at the $N+1-r$th stage,
comes with the factor $\rho^{-N -2 + r }$, the $2r$
 further derivatives will bring its $\rho^{-1}$-order up
to $\rho^{-N - 2}.$  Hence no other factors of $\rho^{-1}$ 
could have fallen on this factor.  Therefore in the repeated multiplications
by $[h + L^2\rho^{-2} h_{-1} + \dots]$, only the 
repeated choice of the term $h$, of order 0 in $\rho^{-1}$, can have
 given rise to the term  $h_{-N - 2 + r}^{(2r)}$ .  It follows that it retains
its original $L$-order, namely $L^{N + 3 - r}.$ \qed

\subsection{ Completion of Proof of Main Lemma}

Our purpose is now to show that by using the joint $\rho, L \rightarrow \infty$
asymptotics one can recover the complete Taylor expansions of all the $h_j$'s
from the wave invariants of the meridian torus and its iterates. That is, to
complete the
\bigskip

\noindent{\bf Proof of the Main Lemma}: We will prove by induction that from
the wave invariants $a_{T_{pL}, -\half + k}$ with $k \leq N$ and all $p \in {\bf N}$,
we can determine the $2k + 4$ -jet of $h$, the $2k$-jet of $h_{-1}$,..., the
$2k + 2 - 2n$-jet of $h_{-n}$ at $\xi = 0.$ (for $n \leq k+1.$)

We note that, unlike in the non-degenerate case,  the principal wave invariant determines the 2-jet of $h$, and
gives no new information under iteration.  We therefore begin the induction
 with the subprincipal wave invariant $a_{T, -\half}$.

From the explicit formula above for the subprincipal wave invariant
 $a_{L, -\half}$ it
is evident that   the 2-jet term in $h$ is old information, 
while the other two terms
differ in the power of
$L$.  Hence they decouple under iteration $L \rightarrow p L$ and we can determine
the 4-jet of $h$ and the 0-jet of $h_{-1}$ from the first two wave invariants. Hence
the induction hypothesis is correct at the first stage.

Assume the induction hypothesis for the $k-1$st stage.  Then the only new information at the kth stage
is that contained in the terms $h_{-k-1}(0)$, $h_{-k}^{(2)}(0), \dots, h^{(2k + 4)}(0)$.  By the proposition
above, they occur linearly in the monomials containing them and the monomials have also the factor of $L$
to the powers $L^{k+1}, L^k, \dots.$  Hence the terms containing the new data decouple as $p \rightarrow \infty$,
and the new data can be determined as stated. This completes the inductive argument.

It follows that we can determine the full Taylor expansions of the $h_{j}$'s  at $\xi = 0.$  Since they
are real analytic they are completely determined.  Then from the homogeneity of $H_j$, we can determine
$H_j$ from $h_j$ and hence the entire function $\hat{H}$ is determined. \qed
\bigskip

\noindent{\bf Remark} In the above argument, we are able to drop the terms involving only previously known
data because of the universal nature of the  polynomials $P_k$.  This universality depends on the fact that
we are only comparing wave data for quantum torus integrable Laplacians.

\section{Proofs of Corollaries 1 and 2}

We now show that the symplectic equivalence class of a metric in ${\cal R}^*$ is spectrally determined among
metrics in this class.  
\bigskip

\noindent{\bf (5.1)~~ Proof of Corollary 1} By the Main Lemma, the principal symbol
$H_g(I_1, I_2)$ as a function of the action variables is spectrally determined for a
metric $g$ in ${\cal R}^*.$  By Propositions (1.5.2-3), $H_g$ determines the
geodesic flow up to symplectic equivalence among metrics in ${\cal R}^*$.
\qed
\bigskip

\noindent {\bf Remark}  The Corollary could also be proved by noting that the first return time $\tau_E$
is spectrally determined.  But geodesic flows of in the class ${\cal R}^*$ are symplectically equivalent
if and only if they have the same first return times $\tau_E$.  (See also [C.K] for an equivalent statement.)
Also, it should be noted that the first return time could be determined from the wave invariants at the
equator; hence Corollary 1 would also follow from Guillemin's inverse result for non-degenerate elliptic
closed geodesics.
\bigskip

\noindent{\bf (5.2)~~ Proof of Corollary 2}: Since $\tau_E(I)$ is spectrally
determined, so is the function
$s(i(I))$ of Proposition 1.3.6.  It is given by
$$s(i(I)) = \int_{i(I)}^{\pi - i(I)} f(cos u )  sin (u) (sin^2 u - sin^2
(i(I)))^{-\half}_{+} du $$
 or, putting $x = cos u$ for $u \in [0, \frac{\pi}{2}]$ and $-x =
cos u $  for $u \in [\frac{\pi}{2}, \pi]$, by
$$s(i(I)) = \int_0^{cos(i(I))} [f(x) + f(-x)] (cos(i(I))^2 - x^2)_+^{-\half} dx.$$

It therefore suffices to show that 
$$T f(u) = \int_0^u [f(x) + f(-x)] (u^2 - x^2 )_+^{-\half} dx$$
determines $ [f(x) + f(-x)]$. But $[f(x) + f(-x)]$ is smooth and even 
so may be written as $g(x^2)$ for a smooth $g$; changing variables
$y = x^2$, and $v = u^2$ we get 
$$T f(v) =  \int_o^1 g(y) y^{-\half} (y - v)_+^{-\half}  dy.$$
Thus $T$ is a standard Abel transform, which is well-known to be invertible. It
follows that $g(y) y^{-\half}$ is spectrally determined;  hence so is the even part
of $f$. \qed
 \bigskip

\section{Proof of Final Lemma}

To complete the proof of the Theorem, we need to show that $\hat{H}$ determines a metric in ${\cal R}.$
It is plausible that this can be done for the following reason: The spectrum of $\Delta$ is the
set of values $\{\hat{H}(n, k+ \half)^2\}$ of $\hat{H}^2$ on the integer points of the
action cone $\Gamma.$ On the other hand since $\hat{I}_1 = \frac{\partial}{\partial
\theta}$, the set of values $\hat{H}(n, k + \half)$ for fixed $n$, is just the
spectrum $\{E_{nk}\}$ of the the singular Sturm-Liouville operator
$$L_n = -(\frac{d}{dr})^2 + q_n(r), \;\;\;\;\;\;\;\;\;\;\;q_n(r):= q(r) + \frac{n^2}{a(r)^2},\;\;\;\;\; q(r) = -
\frac{2 a(r)a''(r) - (a'(r))^2}{ a(r)^2}$$
obtained by separating variables in $\Delta$, fixing the eigenvalue of
$\frac{\partial}{i \partial \theta}$ equal to $n$, and putting the radial part
in normal form.
  Hence,  from the coefficients of  $\hat{H}$ we can determine  $Spec (L_n)$ for each
$n$.  That is, from $Spec(\Delta)$ we have determined the {\it joint} spectrum of
$(\Delta, \frac{\partial}{\partial \theta}).$  It would thus remain to show that the
metric can be determined from this joint spectrum, a much more elementary inverse
result which has been stated several times in the literature ([Kac],[B]
[Gur]). Since these prior
discussions seem to us somewhat sketchy and incomplete, we  give a
self-contained proof below which was found before we were aware of these references.
\bigskip

\noindent{\bf ~~ Proof of Final Lemma}
\bigskip

The proof is basically to write down explicit expressions for $H$ and $H_{-1}$
in terms of the metric (i.e. in terms of $a(r)$) and then to invert the expressions
to determine $a(r)$.  
The first step is therefore to construct an initial part of the quantum normal form
explicitly from the metric.  Up to now, we only know that a polyhomogeneous normal
form exists.

To be sure, the principal term $H$ has already been implicitly expressed in terms
of the metric: Knowledge of $H$ is equivalent to knowledge of the level set $\{H =
1\}$ and hence to knowledge of the function $F(I_1)$ in (1.3.1).  Unfortunately, we
have seen in
Corollary (5.2) that $H$ only determines the `even part' of the metric.  Hence we
need to determine at least one of the subprincipal terms $H_{-j}$.  It turns out
that only $H_{-1}$ is needed in addition to $H$ to determine $g$.  Since the
calculation of $H_{-1}$ requires a new calculation of $H$, we start calculating
both from scratch.

To determine $H$ and $H_{-1}$ in terms of $g$, we are going to study the spectral
asymptotics of $\sqrt{\Delta} =   \hat{H}(\hat{I}_1, \hat{I_2})$ along `rays of
representations'  of the quantum torus action, i.e. along multiples of a given
lattice point $(n_o, k_o).$  Such rays are the quantum analogue of rays
$\Rr^+ T_I \subset T^*S^2-0$ thru invariant tori and are basic to homogeneous
quantization theory [G.S.3].  The basic idea is that the lattice points $(n_o, k_o +
\half)$ parametrize tori $T_{n_o, k_o}$ satisfying the Bohr-Sommerfeld quantization
condition.  To each such quantizable torus one can construct a joint eigenfunction 
$\phi_{n_o, k_o}$ of $(\hat{I}_1, \hat{I}_2)$ by the WKB method. The $\{\phi_{n,k}\}$
are eigenfunctions of $\Delta$ with complete asymptotic expansions along rays. By
studying the eigenvalue problem as $|(n,k)|\rightarrow \infty$ we can determine 
the $H_{-j}$'s.

The WKB method we employ is closely related to the classical WKB method for
constructing quasi-modes (cf. [CV.3] and the Appendix), except that we have
an internal rather than an external Planck constant.  Let us recall the relevant
terminology and notation from the latter case.  For
each torus
$T_I$, we denote by
${\cal O}^{\mu}(T_I, A)$ the space of oscillatory integrals (semi-classical Lagrangean
distributions) associated to $T_I$, with semi- classical parameters $\{k_m\}$ in
a set $A \subset \Rr^+$ to be specified by the quantization condition. Such
oscillatory integrals have the  form 
$$u(x, k) = k^{ \frac{n}{2} + \mu} \sum_{\ell \in L} \int_{{\cal V}_{\ell}} 
e^{i k \psi_{\ell}(x, \xi)}
\alpha(x, \xi, k) d\xi \leqno(6.1)$$
where the projection of $T_I$ is covered by open sets ${\cal U}_{\ell}$,
 where $T^*({\cal U}_{\ell}) \simeq
{\cal U}_{\ell} \times {\cal V}_{\ell}$ and where the phase $\phi_{\ell}(x, \xi),$ 
with $(x, \xi) \in {\cal U}_{\ell} \times {\cal V}_{\ell}$, parametrizes a part of
$T_I$. The amplitude is a classical symbol in $k$ of order $\mu$. For
further details, we refer to [C.V.3, \S 8].

As discussed above,  the torii $T_I$ project $2 - 1$ to the  annuli $r_{+}(I) \leq r
\leq
r_{-}(I)$ in
$S^2$ and have fold singularities along the extremal parallels. Away from 
the parallels, an associated
quasi-mode is given by a sum of two simple WKB functions
 $\alpha_{\pm}(r, \theta) (x) e^{i k s \psi_{\pm} (r,
\theta)}.$   Since the actual $\Delta$- eigenfunctions $\phi_{n_o, k_o}$ are 
quasi-modes attached to the Bohr-Sommerfeld torii $T_{n_ok_o}$,  they have such a form
modulo
 $|(n_o, k_o)|^{-\infty}.$  And, since $\phi_{n_o, k_o}$ is an
 exact $\frac{\partial}{\partial \theta}$-eigenfunction, 
its phases must take the form $\phi_{\ell} (r, \theta) = n_o
\theta + S_{n_o k_o}(r)$ with $\theta$-independent amplitudes 
 in polar coordinates. It follows that   
 $\phi_{n_o, k_o }(r, \theta)$ has the form  
$e^{i n_o \theta} 
f_{(n_o,k_o)}(r)$, where $f_{(n_o,k_o )}(r)$ is
an oscillatory integral in the $r$-variable.  It is of course associated to the
pushed-forward Lagrangean $\Lambda_{n_o k_o}:= p_* T_{n_o k_o}$ where $p : T^*S^2
\rightarrow T^*[0, L]$ is the projection induced from the map $(r, \theta)
\rightarrow r.$ In the case of the meridian torii $T_{0, k_o}$, the pushforward is
just a horizontal line
$p_r = C$ in
$T^*[0,L]$.  In the other cases, the $\Lambda_{n_o k_o}$ is a
 closed curve projecting to an interval $[r_{+}(n_o k_o), r_{-}(n_o k_o)]$
with fold singularities at the turning points  (endpoints).  The curve is given by an
equation of the form
$H_{n_o} (r, p_r) = E$ where $H_{n_o}(r) =  p_r^2 - \frac{n_o^2}{a(r)^2}$ is the
radial Hamiltonian and where $E$ was the level of the torus.

The  radial part $f_{n_o,k_o}(r)$ of $\phi_{n_o, k_o}$,
  is an eigenfunction of the radial operator
$D_{n_o} = -(\frac{d}{dr})^2  - \frac{a'}{a} \frac{d}{dr} + \frac{n_o^2}{ 
a(r)^2}$ arising from separating variables in $\Delta$.  Before proceeding,
we simplify by conjugating $D_{n_o}$ to the 1/2-density radial Laplacian
$$\hat{D}_{n_o} :=a(r)^{\half} D_{n_o} a(r)^{-
\half} = D_r^2 + \frac{n_o^2}{a(r)^2} + W$$
where $W = a(r)^{\half} [-(\frac{d}{dr})^2 - \frac{a'}{a} \frac{d}{dr}]
a(r)^{ - \half}.$ 
(Note that the volume form $d v_g = a(r) dr
d\theta$ of $(S^2,g)$ projects to $a(r) dr.$)  Thus, we   view the radial
eigenfunction as having the form $g_{n_o,k_o}(r)
\sqrt{a dr}$ and  apply the WKB method to the eigenvalue problem
$$ \hat{D}_{n_o} g_{n_o,k_o}(r) \sim (H(n_o,k_o + \half) + H_{-1} (n_o,k_o + \half)
+ \dots)^2 g_{n_o,k_o}(r). \leqno (6.2)$$

 Although the coefficients become
singular at
$r = 0, L$, the standard WKB theory applies in the interior because it applies to
quasi-modes of $\Delta$ on $S^2$. 
Away from the turning points, the
radial part of the 1/2-density eigenfunction therefore has the form 
$$g_{n_o,k_o} = \sum_{\pm}  [e^{ \pm  i 
S_{n_o k_o}(r) } \sum_{m=0}^{\infty} \alpha_{n_o, k_o ; m} (r) ]\leqno(6.3)$$
where the phase $S_{n_o, k_o}$ is homogeneous of degree 1 
and where the amplitude $\alpha_{n_o, k_o ; m}$ is homogeneous of degree $- m$
in $(n_o, k_o + \half).$  These homogeneities replace the powers of $k$ in the
non-homogeneous theory described above.
 
The Bohr-Sommerfeld quantization condition on $T_I$ is that:
$$ \frac{1}{2\pi} I = (n_o,k_o) + \frac{1}{4} \mu_o,\;\;\;\;\;\;(n, k) \in \Gamma
\cap \Z^2$$
where $\mu_0 = (0, 2)$ is  the Maslov index (cf. [CV.3, \S 4]).  It is satisfied
by $T_{n_o, k_o}$ and hence by
the radial Lagrangean $\Lambda_{n_o, k_o}$. The local WKB ansatz (6.3) therefore
extends to a quasi-mode of infinite order associated to the global  $\Lambda_{n_o,
k_o}$. Our purpose now is to write down and solve the first two transport equations,
which are
 needed to determine
$H$ and $H_{-1}$.  For background on the relevant aspects of the WKB method, we refer
to the Appendix.

The transport equations are obtained by separating out terms of like order in
the asymptotic eigenvalue problem
$$\hat{D}_{n_o} \sum_{\pm}  [e^{ \pm  
S_{n_o k_o}(r) } \sum_{m=0}^{\infty} \alpha_{n_o, k_o ; m} (r) ] \sim (H(n_o, k_o + \half) + H_{-1}
(n_o, k_o + \half) + \dots )^2 \sum_{\pm}  [e^{ \pm  
S_{n_o k_o}(r) } \sum_{m=0}^{\infty} \alpha_{n_o, k_o ; m} (r) ].$$
The leading term, of order 2, is the eikonal equation
$$|S_{n_o k_o}'(r)|^2 +
\frac{n_o^2}{a(r)^2} = H (n_o, k_o + \half)^2 \leqno(6.4 a)$$ 
whose solution is the first order phase function
 $$S_{n_o k_o}(r):= \int \sqrt{H(n_o, k_o)^2 -
\frac{n_o^2}{ a(r)^2}}dr. \leqno(6.4b) $$  
The Bohr-Sommerfeld quantization condition on $\Lambda_{(n_o,k_o)}$ thus reads:
$$  I_{n_o} (E) := \mbox{Area}\;\;\;\{H_{n_o} \leq E\} = 2\pi   (k_o
+ \half)\;\;\;\; \mbox{with}\;\;\;   E = H(n_o, k_o + \half).$$

The first transport
equation, 
$2 \alpha_{n_o, k_o; 0}' S' + \alpha S' = 0$, is solved by 
  $$\alpha_{(n_o, k_o; 0}(r) =
  [E_{n_o, k_o}
- \frac{1}{ a(r)^2}]^{-\frac {1}{4}} \leqno (6.5)$$
 away from the turning points. Here,
$E_{n_o, k_o} = \frac{H(n_o, k_o +\half)^2}{n_o^2}$.  The solution is of course
determined only up to a constant, and we have normalized it so that $\alpha_{(n_o,
k_o; 0}$ is homogeneous of order 1.

 In the usual way (see the Appendix), we will interpret $\alpha_{n_o, k_o; 0}$ as
the coefficient of the $\Xi_{H_{n_o}}$- invariant  1/2-density 
$$\nu_0 =   [E_{n_o, k_o}
- \frac{1}{ a(r)^2}]^{-\frac {1}{4}} \sqrt{dr} $$ on 
$\Lambda_{n_o k_o} = \{
H_{n_o}(r, p_r) = H( n_o, k_o + \half)\},$  where $\Xi_{H_{n_o}}$ is the Hamilton
vector field  of
$H_{n_o}$. We then re-write the higher coefficients 
 $\alpha_{n_o, k_o ; m} (r)$ in the form $\alpha_{n_o, k_o; m} (r) \nu_0.$ 

The second transport equation (of order zero) then has the form:
$$\Xi_{H_{n_o}} \alpha_{n_o, k_o, -1} = -   \alpha_{n_o, k_o, 0}^{-1}
(\frac{d}{dr})^2 
\alpha_{n_o, k_o, 0} + W  + 
2 H(n_o, k_o + \half) H_{-1} (n_o, k_o + \half),\leqno(6.6)$$ where $r$ denotes the
local coordinate on $\Lambda_{n_o k_o}$ obtained by pulling back the base coordinate
under the projection.  The integral of the left hand side over the closed curve
$\Lambda_{n_o k_o} = \{H_{n_o} = H(n_o, k_o + \half)\}$ with respect to the
$\Xi_{H_{n_o}}$-invariant density 
$$\alpha_{n_o, k_o, 0}^2 dr = \frac{1}{\sqrt{E_{n_o,
k_o} - \frac{1}{a(r)^2}}} dr$$ must equal zero.  This gives a formula for the
first correction to the Bohr-Sommerfeld eigenvalue $H(k_o, n_o)$: 
$$H_{-1} (n_o, k_o + \half) = \frac{1}{2 T (n_o, k_o) H(n_o, k_o + \half)} 
\int_{r_{-}(n_o, k_o)}^{r_{+}(n_o, k_o)} [   \alpha_{n_o, k_o, 0} (\frac{d}{dr})^2 
\alpha_{n_o, k_o, 0} - \frac{1}{2} W \alpha_{n_o, k_o, 0}^2] dr . 
\leqno(6.7)$$
 Here,  $T(n_o,
k_o)$ denotes the period of the $\Xi_{H_{n_o}}$ flow on $\Lambda_{n_o k_o}$ and
$r_{\pm}(n_o, k_o)$ are the turning points.  Plugging in (6.5) we get, formally,
the expression
$$H_{-1} (n_o, k_o + \half) = \frac{1}{2 T (n_o, k_o) H(n_o, k_o + \half)} 
\int_{r_{-}(n_o, k_o)}^{r_{+}(n_o, k_o)}   (E_{n_o, k_o} -
\frac{1}{a(r)^2})^{-1/4} (\frac{d}{dr})^2 (E_{n_o, k_o} -
\frac{1}{a(r)^2})^{-1/4} dr \leqno(6.8)$$
$$ - \frac{1}{2 T (n_o, k_o) H(n_o, k_o + \half)} 
\int_{r_{-}(n_o, k_o)}^{r_{+}(n_o, k_o)} \frac{W}{\sqrt{E_{n_o, k_o} -
\frac{1}{a(r)^2}}} dr.$$

Actually, the integral in (6.8) diverges at the turning points, and the correct
formula is the regularization obtained (for instance) by the  method of the Maslov
canonical operator. For the sake of completeness, we provide   an exposition of this
method in the appendix. Roughly speaking, it regularizes (6.8) by formally
integrating the $\frac{d}{dr}$ derivatives by parts and by moving them outside the
integral (in the appropriate way)  as derivatives in the energy level $E$. A crucial
consequence of the regularization is that only first derivatives of $a$ appear in
the formulae for $H_{-1}(n_o, k_o + \half).$

Thus the first term of (6.8) is regularized by  
$$\frac{C_1 }{ T (n_o, k_o) H(n_o, k_o + \half)}  
\partial_E^2  
\int_{ r_{-}(n_o, k_o)}^{r_+(n_o, k_o)} \frac{a'(r)^2}{a(r)^6} (E -
\frac{1}{a(r)^2})^{-\half} dr|_{E = E_{(n_o, k_o)}}.\leqno(6.8.1 reg)$$
Here and below $C_i$ denote (non-zero)  constants which we will not need to determine.
For the second term of (6.8), we note that
$$\int_{ r_{-}(n_o, k_o)}^{r_+(n_o, k_o)} W \alpha_{n_o, k_o; 0}^2 =
 \int_{ r_{-}(n_o, k_o)}^{r_+(n_o, k_o)}
\frac{d}{dr} (\alpha_{n_o, k_o; 0}^2 a^{\half}) \frac{d}{dr} a^{-\half} dr  + 
\int_{ r_{-}(n_o, k_o)}^{r_+(n_o, k_o)}
\alpha_{n_o, k_o; 0}^2 a^{\half} \frac{a'}{a} \frac{d}{dr} a^{-\half} dr. $$
After some simplification, this regularizes to: 
$$  \frac{C_2 }{ T (n_o, k_o) H(n_o, k_o + \half)} \partial_E
\int_{r_{-}(n_o, k_o)}^{r_{+}(n_o, k_o)} \frac{a'(r)^2}{a(r)^4}(E -
\frac{1}{a(r)^2})^{- \half} dr|_{E = E_{(n_o, k_o)}} \leqno(6.8.2 reg)$$
$$+  \frac{C_3}{  T (n_o, k_o) H(n_o, k_o + \half)} 
\int_{r_{-}(n_o, k_o)}^{r_{+}(n_o, k_o)} \frac{a'(r)^2}{a(r)^2} (E -
\frac{1}{a(r)^2})^{- \half} dr|_{E = E_{(n_o, k_o)}}.$$

Now let us return to the inverse problem.  We begin from the fact that $\hat{H}(n_o,
\hat{I}_2)$ is a known  function of the  variable $I:=\hat{I}_2$ for each $n_o$. 
Its principal symbol $H_{n_o}(I):= H_1 (n_o, I)$ is then a known function and hence
the inverse function 
$$I_{n_o} (E) = \int_{r_{-}(E)}^{r_+ (E)} \sqrt{E - \frac{1}{a(r)^2}} dr$$
 satisfying $H_{n_o} (I_{n_o}(E)) = E$ is a known of $E$.  This of course
presupposes that $\partial_I H_{n_o}(I) \not= 0$, which follows from the
non-degeneracy assumption (1.1.6). We
may write the integral in the form
$$ \int_{\Rr} ( E - x)_+^{\half} d\mu(x)$$
where $\mu$ is the distribution function of $\frac{1}{a^2}$, i.e.
$$\mu (x):= |\{r: \frac{1}{a(r)^2} \leq x\}|$$
with $|\cdot|$ the Lebesgue measure.  The above integral
is an Abel transform and as mentioned above it is invertible.  Hence 
$$d \mu(x) = \sum_{r: \frac{1}{a(r)^2} = x} |\frac{d}{dr} \frac{1}{a(r)^2} |^{-1}
dx$$  is a spectral invariant.  Some simplification leads to the conclusion that the
function
$$ J(x) := \sum_{r: a(r) = x} \frac{1}{|a'(r)|}$$
is known from the spectrum.  By the simplicity assumption on $a$, there are just two
solutions of
$a(r) = x$; the smaller will be written
$r_{-}(x)$ and the larger,
$r_{+}(x).$ Thus,  the function
$$J(x) = \frac{1}{|a'(r_{-}(x))|} + \frac{1}{|a'(r_{+}(x))|} \leqno(6.9)$$
is a spectral invariant.

Now let us turn to the $H_{-1}$ expression.  Since $H(I_1, I_2)$ is a spectral
invariant, the factors $H(n_o, k_o + \half)$ and $T(n_o, k_o)$ are spectral
invariants.  Hence we may remove them from the expression for $H_{-1}$ and
still get a spectral invariant.  For various universal constants $C_1,C_2, C_3$
it takes the form
$$[C_1 \partial_E^2 \int \frac{( a')^2}{a^6} ( E  -
\frac{1}{a^2})_{+}^{-\frac{1}{2}} dr + C_2 \partial_E  \int \frac{(a')^2}{a^4}
(E -  \frac{1}{a^2} )_{+}^{- \half} dr \leqno(6.10)$$
$$+ C_3 \int \frac{(a')^2}{a^2}
(E -  \frac{1}{a^2} )_{+}^{- \half} dr] |_{E = \frac{H(I_1,I_2)}{I_1^2}}.$$
  By a change of variables, we may rewrite (6.10) in the form  
$$[C_1 \partial_E^2  \int K(x) x^{\frac{3}{2}} (E - x)_+^{-\half} dx
+  C_2 \partial_E \int K(x) x^{\half} (E - x)_+^{-\half} dx \leqno(6.11)$$
$$ + C_3   \int K(x) x^{-\half} (E - x)_+^{-\half} dx]|_{E =
\frac{H(I_1,I_2)}{I_1^2}}$$
 where
$$K(x) = |a'(r_{-}(x))| + |a'(r_{+}(x))|.\leqno (6.12)$$
All values of $E$ which occur as ratios $ \frac{H(I_1,I_2)}{I_1^2}$ give
spectral invariants, so (1.10) (as a function of the variable $E$) is a
spectral invariant. 

We now claim that $K$ itself is a spectral invariant.  To determine it from (6.11)
we rewrite (6.11) in terms of the fractional integral operators (cf. [G.Sh, Ch.1 \S
5.5]) 
$$I_{\alpha} f(E) = f * \frac{x_+^{\alpha - 1}}{\Gamma(\alpha)} (E) = 
\frac{1}{\Gamma( \alpha)}
\int_0^E f(y) (E - y)^{\alpha - 1} dy$$
on the half-line $[0, \infty].$
 These operators satisfy
 $$I_{\alpha} \circ I_{\beta} = I_{\alpha + \beta},\;\;\;\;\; I_{-k} =
(\frac{d}{dx})^k.$$ 
 Thus (6.11) equals ${\cal L}(K)$ where
${\cal L}$ is the fractional integral operator  
 $$ {\cal L}:= C_1 I_{-3/2} x^{\frac{3}{2}} + C_2 I_{-\half} x^{\half}  + C_3 
I_{\half} x^{-\half}. \leqno(6.13)$$
To solve for $K$ we apply
$I_{-\half}$ to ${\cal L} K $
to get
$$C_1' \frac{d}{dx}^2 (x^{\frac{3}{2}} K(x)) + C_2' \frac{d}{dx} (x^{\half} K) + C_3'
x^{-\half} K = I_{-\half} {\cal L} K.\leqno(6.14)$$
This equation determines $K$ up to a solution $f$ of the associated homogeneous
equation, essentially an Euler equation.   But also $K = 0$ on $[0, a(r_o)^{-2}]$
Since no homogeneous solution can have this property, $K$ is uniquely determined
by this boundary condition.

It follows that both (6.9) and (6.12) are spectral invariants.  But from
 $a + b$ and $\frac{1}{a} + \frac{1}{b}$ one can determine the pair $(a,b)$.  Hence
 $a'(r)$ is determined from the spectrum.  Since $a(0)=0$ this determines
$a$ and hence the surface. \qed

\section{Appendix}

The purpose of this appendix is to give an algorithm for calculating the higher
order terms in the quasi-classical approximation of eigenvalues for 1 D Schrodinger
operators  $-\frac{h^2}{2} \frac{d^2}{dx^2} + V$
 with confining potentials.  In particular, we carry out the calculation
of the $h^2 E_n^{(2)}$ term, which was used in the proof of the Final Lemma.
  
The algorithm is based on the Maslov method of canonical operators. Expositions
and refinements of this method can be found, among other places, in Maslov's book [M],
in the article of  Colin de Verdiere [CV.3] (and in its references), and in the recent
book of Bates- Weinstein [B.W].  Although these references explain the construction of
the canonical operator and prove the existence of complete quasi-classical eigenvalue
expansions, they do not generally go on to describe the calculation of the terms.
An exception is the original book of Maslov [M], which does calculate the first
two or three terms; but the method of canonical operators is abandoned
at this point in favor of some methods of special functions.   As we will show, the
canonical operator method gives the required corrections quite efficiently.  
\bigskip

\noindent{\bf The set-up}
\bigskip

We are concerned with the semi-classical eigenvalue problem:
\bigskip

$$\left\{ \begin{array}{l} [-\frac{h^2}{2} \frac{d^2}{dx^2} + V] \psi_n (x,h) =
E_n(h) \psi_n (x,h)\\
\langle \psi_n, \psi_m \rangle = \delta_{mn} + O(h^{\infty})\\
E_n(h)  = E_n^{(1)} (h) + h^2 E_n^{(2)} + h^3 E_n^{(3)} + \dots \end{array}
\right.$$
\bigskip

The unknown function $\psi_n (x,h)$ is an oscillatory associated to a Lagrangean
of the form $\Lambda_n := \{H = E_n^{(1)}(h)\}$ where
$E_n^{(1)}(h)$ is determined by the Bohr-Sommerfeld-Maslov 
quantization condition:
$$\frac{1}{2 \pi h} \int_{\Lambda_n} \xi \cdot dx = n + \frac{1}{4} \mu.$$
Here, $\mu$ is the Maslov index of $\Lambda_n$; it equals $2$ for connected level
sets of Hamiltonians of the form $H(x,\xi) = \xi^2 + V(x).$

To find the higher order corrections $E_n^{(k)}$ we consider the Maslov canonical
operator 
$${\cal U}_h : S^m (\Lambda_n, \Omega_{\half} \otimes {\cal M}) \rightarrow
{\cal O}^m (\Rr, \Lambda_n).$$
We follow here the notation and terminology of [B.W][CV.3]: $S^m (\Lambda_n,
\Omega_{\half}
\otimes {\cal M})$ is the space of symbolic sections of the bundle of 1/2-densities
times Maslov factors and ${\cal O}^m (\Rr, \Lambda_n)$ is the space of oscillatory
integrals associated to $\Lambda_n$.  There is a natural symbol map in the reverse
direction; any inverse modulo $O(h)$ is a quantization or canonical
operator.  Its existence is equivalent to the condition that $\Lambda_n$ satisfy
the BSM quantization condition.  For background we again refer  to [CV.3][B.W].

We also denote by $\Xi_H$ the Hamilton vector field of $H$, by ${\cal L}_{\Xi_H}$
the Lie derivative on any bundle, by $s$ a nowhere vanishing section of ${\cal M}$,
and by $\rho$ a
${\cal L}_{\Xi_H}$-invariant density on $\Lambda_n$ (for a fixed $n$). 
By surjectivity of ${\cal U}_h$, 
we can 
write an oscillatory integral associated to $\Lambda_n$ in the  form
$$\psi_n (x, h) |dx|^{\half} \sim {\cal U}_h [ e^{\frac{i}{h} \phi} \cdot
\sum_{j=0}^{\infty} h^j f_j
\rho^{\half} \otimes s].$$
Our aim is to determine the quasi-classical series $\sum E_n^{(j)} h^j$  and
coefficient functions
$f_j$ for which the asymptotic eigenvalue problem is solvable.

We begin by constructing local solutions.  Thus we first
consider $x$-projectible pieces of  $\Lambda_n \subset T^*\Rr$: pieces which projects
regularly from a  neighborhood of $\lambda \in \Lambda_n$ to a neighborhood of $x \in 
\Rr.$  Restricted to densities supported on such pieces, the Maslov canonical operator
is truly canonical: if
$S(x)$  is a phase locally parametrizing $\Lambda_n$, then 
$${\cal U}_h [ e^{\frac{i}{h} \phi} \cdot
\sum_{j=0}^{\infty} h^j f_j
\rho^{\half} \otimes s] =  e^{\frac{i}{h} S} \cdot
\sum_{j=0}^{\infty} h^j a_j$$
for some smooth coefficients $a_j$.  We may then substitute this expression into
the eigenvalue problem and obtain eikonal and transport equations. The eikonal
equation $(S')^2 + V(x) = E_n^{(1)} (h) = 0$ has been solved by our choice of
phase, so the transport equations become:
$$\left\{ \begin{array}{l} a_o  \frac{d^2}{dx^2}S + 2 \nabla a_o \nabla S = 0 \\
a_1  \frac{d^2}{dx^2} S + 2 \nabla a_1 \cdot \nabla S - i  \frac{d^2}{dx^2} a_o = 2 i
E_n^{(2)} a_o \\ a_2  \frac{d^2}{dx^2} S + 2 \nabla a_2 \cdot \nabla S - i 
\frac{d^2}{dx^2} a_1 = 2 i [E_n^{(3)} a_o + E_n^{(2)} a_1 ] \end{array} \right. $$
and so on.  As is well-known (cf. e.g. [B.W]), these equations may be put into 
geometric form by observing that $\nabla S \cdot \nabla$ is the projection to
$\Rr$ of ${\cal L}_{\Xi_H}$ and that $[a  \frac{d^2}{dx^2} S + 2 \nabla \cdot S] |dx|
= div (a^2 \nabla S) |dx| = {\cal L}_{\Xi_H} (a^2 |dx|).$ Hence the transport
equations become 
$$\left\{ \begin{array}{l} {\cal L}_{\Xi_H} (a_o |dx|^{\half}) = 0 \\
{\cal L}_{\Xi_H} (a_1 |dx|^{\half}) = (2 i E_n^{(2)} a_o + i  \frac{d^2}{dx^2}
a_o)|dx|^{\half}
\\ {\cal L}_{\Xi_H} (a_2 |dx|^{\half}) = 2 i [(E_n^{(3)} a_o
+ E_n^{(2)} a_1) + i  \frac{d^2}{dx^2} a_1)] |dx|^{\half} \end{array} \right. $$
The invariant 1/2-density is given by the well-known formula $a_o = (E -
V)^{-\frac{1}{4}}$, or equivalently by $\frac{|dx|^{\half}}{p^{\half}}$ on
$\Lambda_n.$  If we write $a_j |dx|^{\half} = f_j \rho^{\half}$ then $f_0 = 1,
f_j = \frac{a_j}{a_o}$ and the transport equations become
$$\left\{ \begin{array}{l} 
\Xi_H f_1  = (2 i E_n^{(2)} + i a_o^{-1}  \frac{d^2}{dx^2} a_o)
\\ \Xi_H f_2  = 2 i [(E_n^{(3)}
+ E_n^{(2)} \frac{a_1}{a_o}) + i a_o^{-1} \frac{d^2}{dx^2} a_1)]  \end{array} \right.
$$ Here, the expressions in $a_o, a_1,\dots$ are understood to have been lifted up
to $\Lambda_n$.  

The eigenvalue corrections $E_n^{(k)}$ are determined by integrating both sides
of the transport equations over the level $\{H = E_n^{(1)}\}$.  Since the equations
are solvable and since the left hand sides will integrate to zero, we get
$$\left\{ \begin{array}{l} 
 E_n^{(2)} = \frac{1}{2i T(E_n^{(1)})} \int_{\{H = E_n^{(1)}\}} [a_o^{-1} 
\frac{d^2}{dx^2} a_o] \rho 
\\ 
E_n^{(3)} = \frac{1}{2i T(E_n^{(1)})} \int_{\{H = E_n^{(1)}\}} 
[ E_n^{(2)} \frac{a_1}{a_o} + i a_o^{-1}  \frac{d^2}{dx^2} a_1] \rho  \end{array}
\right. $$ Here, $T(E)$ is the period of $\Xi_H$ at level $E$. 
Parametrizing $\{H = E_n^{(1)}\}$ as a graph over the $x$-axis away from the turning
points, the invariant density takes the form $\frac{dx}{\sqrt{E - V}}$ with
$E = E^{(1)}_1$.  Hence at least formally the eigenvalue corrections are given by
$$ E_n^{(2)} = \frac{1}{2i T(E)} \int_{x_{-}(E)}^{x_{+}(E)} [a_o^{-1} 
\frac{d^2}{dx^2} a_o]
 \frac{dx}{\sqrt{E - V}} =  \frac{1}{2i T(E)} \int_{x_{-}(E)}^{x_{+}(E)}
 (E - V(x))^{-\frac{1}{4}} \frac{d^2}{dx^2} (E - V(x))^{-\frac{1}{4}} dx.$$ 

Unfortunately the integral is ill-defined due to the singularities at the turning
points. The problem is that the Maslov operator cannot be defined  near these points
as a simple pull-back operator.  Rather it should be defined as the composition of
the Fourier transform with the pull-back operator defined over the $\xi$-projection.
This problem and its solution constitute a key aspect of the Maslov method (in
one dimension); we refer to [CV.3][B.W] for extended discussions 

The point which is important for us is that the Maslov method gives a regularization
of the divergent integral. It works in the following way: we introduce a cut-off
$\psi_{\delta}$ supported away from a $\delta$-neighborhood of  the turning points
$x_{\pm}(E).$  More precisely we define $\psi_{\delta}^{\pm}$ on $\Lambda_n$, equal to
one on $(2 \delta, \half T(E) - 2\delta)$  resp. $ (\half T(E) + 2 \delta, T(E) - 2
\delta)$ and equal to zero on $(T(E) - \delta, \delta)$ resp. $(\half T(E) - \delta, 
\half T(E) + \delta).$  We then put
$${\cal U}_h (f \rho \otimes s e^{\frac{i}{h} \phi}) := I_h (\psi_{\delta}
f \rho \otimes s e^{\frac{i}{h} \phi}) + J_h ((1 - \psi_{\delta}) f \rho \otimes s
e^{\frac{i}{h} \phi})$$
where $I_h$ is the pull-back to $\Rr$ under the phase parametrization by $\xi = 
S'(x)$ and where $J_h$ is the Fourier transform of the $\xi$-parametrization.
The notation $\psi_{\delta}$ stands for $\psi_{\delta}^{\pm}$.  For details on
$I_h, J_h$, see [B.W].

Returning to the previous calculation of eigenvalue corrections, we see that what
is missing is the cut-off $\psi_{\delta}$ in the integrals and the contributions
from $J_h$.  We wish to avoid confronting the latter.  Fortunately, it is not
necessary to do so: the fact that the eigenvalues are independent of $\delta$ allows
us to determine the $J_h$ (i.e. the turning point) contribution indirectly.  To see
this, we substitute the cut-off into the formula for $E_n^{(2)}$ to get
$$ E_n^{(2)} =  \frac{1}{2i T(E)} \int_{x_{-}(E)}^{x_{+}(E)}
 (E - V(x))^{-\frac{1}{4}} \frac{d^2}{dx^2} [(E - V(x))^{-\frac{1}{4}} \psi_{\delta}]
dx + II_{\delta}$$
with $II_{\delta}$ the contribution from $J_h$.  Since the integral is now nicely
convergent we can integrate by parts and simplify to the form
$$\frac{1}{16 T(E)} \int_{x_{-}(E)}^{x_{+}(E)} \frac{V'(x)^2}{(E - V)^{\frac{5}{2}}}
\psi_{\delta} dx -
\frac{1}{ T(E)} \int_{x_{-}(E)}^{x_{+}(E)} \frac{V'(x)}{(E - V)^{\frac{3}{2}}}
\psi_{\delta}'(x) dx. $$
The first term tends to
$$\frac{1}{12 T(E)} \frac{d^2}{dE^2} \int_{x_{-}(E)}^{x_{+}(E)} \frac{V'(x)^2}{(E -
V)^{\frac{1}{2}}}
\psi_{\delta} dx$$
as $\delta \rightarrow 0$. The second expression only involves the Taylor expansion
of $V$ near the turning points.  Its singular part must be cancelled by the singular
part of $II_{\delta}$, leaving a possible `residue' as $\delta \rightarrow 0.$
We claim that this residue is zero: in fact, this is known to be the case since the
first term
is well-defined, independent of $\delta$, and agrees with the  formula given in [M]. 
To give an independent proof that it vanishes, without analysing the $J_h$-term in
detail, we observe that the limit contribution involves only the 2-jet of $V$  at the
turning points. Hence it must agree with the corresponding expression for a harmonic
oscillator at its turning points.  But no such correction occurs.

\end{document}